\documentclass[11pt,twoside]{article} 
\usepackage{times,fancyhdr}
\usepackage{cite}
\usepackage{latexsym}
\usepackage{amssymb}
\usepackage{amsmath}
\usepackage{graphicx}
\usepackage{wasysym}
\usepackage{bigfoot}
\usepackage[nottoc]{tocbibind}

\newcommand{\quantt}{\left<T_{\mu\nu}\right>}
\newcommand{\psistate}{\left|\Psi\right>}
\newcommand{\pme}{^{\prime}}
\newcommand{\alf}{\alpha(r)}
\newcommand{\gam}{\gamma(r)}
\newcommand{\sunmass}{M_{{\mbox{\tiny{$\astrosun$}}}}}
\newcommand{\h}{\tilde{g}}

\date{}

\title{Developments in Black Hole Research: Classical, Semi-classical, and Quantum} 
\author{{\small A. DeBenedictis \footnote{adebened@sfu.ca}} \\
\it{\small Pacific Institute for the Mathematical Sciences,} \\
\it{\small Simon Fraser University Site} \\
\it{\small and}\\
\it{\small Department of Physics, Simon Fraser University}\\
\it{\small Burnaby, British Columbia, V5A 1S6, Canada }}
\begin{document} 

\pagestyle{fancy}
\fancyhead{} 
\fancyhead[EC]{A. DeBenedictis}
\fancyhead[EL,OR]{\thepage}
\fancyhead[OC]{Developments in Black Hole Research}
\fancyfoot{} 
\renewcommand\headrulewidth{0.5pt}
\addtolength{\headheight}{2pt} 

\maketitle 
\vspace{-0.5cm}
\noindent
\hrulefill
\vspace{1cm}

\begin{center}
{\bf ABSTRACT}
\end{center} 
\noindent The possible existence of
black holes has fascinated scientists at least since Michell and
Laplace's proposal that a gravitating object could exist from
which light could not escape. In the 20th century, in light of
the general theory of relativity, it became apparent that, were
such objects to exist, their structure would be far richer than
originally imagined. Today, astronomical observations strongly
suggest that either black holes, or objects with similar
properties, not only exist but may well be abundant in our
universe. In light of this, black hole research is now not only
motivated by the fascinating theoretical properties such objects
must possess but also as an attempt to better understand the
universe around us. We review here some selected developments in black hole
research, from a review of its early history to current topics in
black hole physics research. Black holes have been studied at all
levels; classically, semi-classically, and more recently, as an
arena to test predictions of candidate theories of quantum
gravity. We will review here progress and current research at all
these levels as well as discuss some proposed alternatives to
black holes.

\vspace{7mm}

\newpage
\tableofcontents
\newpage

\section{Introduction}
This paper presents some developments in black hole research from its very early history to modern day. Any manuscript undertaking such a task is bound to be incomplete, the subject matter being enormous. What is intended here is a coherent, reasonably self-contained (and relatively brief) article capturing essential features in black hole history and research at the classical, semi-classical, and quantum level. The goal is to give the interested researcher or student an overview of some of the research that has been done and is currently being pursued in all these areas within a single manuscript. The style is more of a survey than an in-depth study and it is hoped the interested reader will find the references useful for further information. Given limited space and time, there are regrettably many, sometimes glaring, omissions and entire fascinating areas of research had to be left out. It was therefore decided that the bulk of the effort go into reviewing a few selected topics in four dimensional black holes within the context of the original general relativity theory of Einstein and Hilbert and their natural extensions into the quantum realm. A sincere apology goes to the authors of works not included here or accidentally missed. Some of the topics are chosen due to their lasting impact in the field as can be seen, for example, by the number of papers appearing on the arXiv related to these topics, and are not necessarily new. It is hoped that this type of review will give researchers in other areas of gravity a brief overview of the phenomena that these recurring topics comprise.

\pagestyle{fancy} \fancyhead{} \fancyhead[EC]{A. DeBenedictis}
\fancyhead[EL,OR]{\thepage} \fancyhead[OC]{Developments in Black
Hole Research: Classical, Semi-classical, and Quantum} \fancyfoot{}
\renewcommand\headrulewidth{0.5pt}

Black hole research has turned from an obscure, almost ignored area of research to one of the most studied segments in gravitational field theory. Today it is common to see more than a few black hole papers appear on the pre-print archive on a daily basis. It seems that the black hole still has many interesting surprises, from the purely classical, to the purely quantum, and everything in between.

The presentation here is done in a somewhat historical perspective. However, the bulk of the results focus on more recent developments as there are a number of excellent books and reviews from a purely historical point of view (\cite{ref:thornebook}, \cite{ref:rovellihist}, \cite{ref:ashhist} and references therein).

In section 2 we give a brief history of black hole research, which dates at least back to the 1780s. In section 3 we present the classical black holes and the fascinating research that accompanies them to this day. In section 4 we look at semi-classical research which also includes a section on Hawking's amazing result of black hole radiation. In section 5 we study results from quantum gravity, primarily the loop approach, which is not to imply that other approaches are not fruitful. In starting to write that section of the manuscript it seemed that with several differing theories, either no justice could be paid to any of them, or else the section had to focus on the research occurring in just one of them. It seems that loop quantum gravity is closest to the spirit of the rest of the paper and has also produced a number of interesting results. 

Of course, black holes would not be nearly as interesting if not for the fact that there is now reason to believe that they may well exist (perhaps even in abundance) in the universe. We therefore focus on current astrophysical black hole research in section 6 (much of the discussion in this section also applies to the possible detection of \emph{primordial} black holes \cite{ref:primord1} - \cite{ref:primord3}, although many would have evaporated by the present era.) This is perhaps the fastest changing area in black hole research and it seems that the activity in this field will not be dying down any time soon. Finally, in section 7 we also discuss some alternative theories of collapse which avoid the formation of a singularity or even the black hole altogether. 

\section{Black holes: a short history}
The idea that a gravitating object could exist from which not even
light could escape seems to date at least back to the work of the
Reverend John Michell in 1783 \cite{ref:michell}. At this time it
was already known that light traveled at a finite speed from
Roemer's studies of Jupiter's moon Io \cite{ref:roemer}. If light
behaved like a particle, with finite speed, why then could it not
be affected by gravity like other objects?

In 1796 Pierre Laplace, apparently unaware of Michell's work postulated exactly the same thing; that a ``dark star''
could exist from which no light would escape \cite{ref:laplace}.
Both Michell and Laplace calculated that an amount of mass
$M$ must be present within a radius $R=\frac{2GM}{c^2}$ in order
for light not to escape from the object. The circumference
corresponding to this radius was called the \emph{critical
circumference}. As is well known, this value is in (surprising)
agreement with the value given by general relativity theory.

Although it was initially believed that these dark stars could be
populous in the universe, they were later considered to be at
best an academic curiosity,  as the size or density such a body
would possess was considered unphysical. Michell originally
calculated that an object with a similar average density to that
of the sun, would need to be approximately 500 times larger in
order to stop light. Put another way, an object with the same
mass as the sun would need to possess a diameter of a mere 20
kilometers.  As well, studies in the 1800's indicated that light
was a wave possessing no mass and therefore it was believed that
it would not be influenced by gravitational effects.

The situation changed in 1915 when Einstein and Hilbert formulated
the now famous field equations of general
relativity \cite{ref:einstgr}, \cite{ref:hilbgr}, which in the
notation of this manuscript are presented as \footnote{In
subsequent sections we shall be utilizing geometrized units where
$G=c\equiv 1$.}:
\begin{equation}
 R_{\mu\nu}-\frac{1}{2}R\,g_{\mu\nu}=\frac{8\pi G}{c^{4}} T_{\mu\nu}\:. \label{eq:einsteq}
\end{equation}

It was not long after the formulation of the field equations that
astrophysicist Karl Schwarzschild came up with an exact solution
which described the gravitational exterior of a perfectly
spherical star \cite{ref:schw}. The Schwarzschild solution is
probably the most famous non-trivial solution of the field
equations and it admits the following well-known line element in
the spherical coordinate chart:
\begin{equation}
 ds^{2}=-\left(1-\frac{2GM}{c^{2}r}\right)\,dt^{2} + \frac{dr^{2}}{\left(1-\frac{2GM}{c^{2}r}\right)}
 + r^{2}\,d\theta^{2} +r^{2}\sin^{2}\theta \, d\phi^{2}\:. \label{eq:schwline}
\end{equation}

Two things are immediately apparent in (\ref{eq:schwline}): (i) A
singularity is present when $r=0$. This singularity was believed
to be of exactly the same nature as the corresponding one in the
Newtonian theory and was of little concern. (ii) There exists a
singularity in the metric when $r=\frac{2GM}{c^2}$, exactly the
same radius value for which an object in Newton's theory yields
an escape velocity equal to $c$. However, even though it was well
known that light \emph{was} affected by gravity in this new
theory, it was still believed that stars whose radius was less
than $\frac{2GM}{c^2}$ were to be considered as pathological and
not existing in nature.

This period of complacency did not last particularly long. Many
scientists at the time began to worry that something seemingly unnatural appeared in the gravitational field theory and appealed to Schwarzschild's constant
density interior solution \cite{ref:schwdensity}. For a constant
density sphere of density $\rho_{0}$ and radius $a$, Einstein's
equations along with the conservation law $T^{\mu\nu}_{\;\;\;;\nu}=0$
yield the following for the (isotropic) pressure:
\begin{equation}
 p=\rho_{0}\left[\frac{\sqrt{1-2Mr^{2}/a^{3}}-\sqrt{1-2M/a}}{3\sqrt{1-2M/a} -\sqrt{1-2Mr^{2}/a^{3}}}\right], \label{eq:constpress}
\end{equation}
with  $r < a$. In this expression it was noted that as $a$
approaches $\frac{9}{4}M$, the central pressure becomes infinite.
Thus, it was argued, systems with smaller radius-to-mass ratios
could not exist in nature. This argument, however, relied on the
use of an unphysical matter model and perhaps could not be
trusted.

Einstein himself was uncomfortable with the fact that the solution
to his field equations admitted such bizarre structures
\cite{ref:thornebook}. He attempted to disprove the
existence of such compact objects by studying the circular orbits
of massive particles in the Schwarzschild space-time. As he made
the orbits smaller and smaller, the particle velocities increased
until they finally reached the speed of light when located at a
radius of 1.5 times the critical radius. The conclusion was that
a spherical object could not exist whose radius was less than
this one since no particle could exceed the speed of light
\cite{ref:einsteinspaper}. This study, though perfectly correct,
neglected radial motions of the particles which are inevitably
present for all particles with geodesic orbital radii of less than 1.5 times the
critical radius.

In the same year as Einstein's paper was published (1939), Robert Oppenheimer and Hartland Snyder performed
an extremely difficult and pioneering calculation. They studied
the gravitational collapse of a spherically symmetric isotropic
dust cloud within full general relativity. Their seminal results
were published in an article entitled \emph{``On Continued
Gravitational Contraction''}\cite{ref:opsnyd}. In this study they
demonstrated how the dust cloud could collapse and how, when
viewed from an external vantage point, the collapse would slow
down and asymptotically halt as the critical radius was reached.
They also showed that, to an observer co-moving with the dust,
the collapse takes place in finite time. Although dust is by no
means a realistic matter field, Oppenheimer and Snyder's
calculations provided the most complete argument of gravitational
condensation at the time.

One of the continuing opponents to black hole formation was
J.~A.~Wheeler. Wheeler and many other scientists at the
time were quite certain that some physical process must intervene
during the collapse, and that the scenario played out by
Oppenheimer and Snyder's idealized calculation would not occur.
One proposal was that the nucleons present in a realistic
structure would, under extreme conditions, radiate away
\cite{ref:thornebook}. Interestingly, this idea turns out to be
partially correct in the paradigm of black hole evaporation.

By the 1960s computers were at the stage where a more realistic
matter model could be utilized in gravitational collapse
calculations. Such a calculation was carried out by Colgate and
his collaborators at Livermore laboratory \cite{ref:maywhite}. These calculations,
although still perfectly spherically symmetric, took into account
now well known nuclear physics processes inside the matter. These
calculations indicated that for a star of mass greater than
approximately two solar masses, collapse into a black hole was
inevitable. Similar calculations were carried out in the Soviet
Union by Zel'dovich and collaborators yielding similar results
\cite{ref:thornebook}. Such simulations along with the discovery of new
coordinate systems to describe black holes 
\cite{ref:fink} \cite{ref:krusk} aided in easing the scientific community's
skepticism about black holes. In fact, it is well known that
Wheeler became a big believer and is the originator of the term
``black hole''.

With the possible formation of black holes now accepted by much of
the scientific community, black hole research saw the birth of
the now famous Hawking-Penrose singularity theorems
\cite{ref:hawkpen}, the no-hair theorem \cite{ref:nohair_price},
and of course, the Kerr solution to the field equations with all
its interesting properties and peculiarities, and the laws of
black hole mechanics.

The laws of black hole mechanics arose from the analysis due to
various talented scientists and were put in their final form by
J.~Bardeen, B.~Carter and S.~Hawking \cite{ref:bhmech}. In a
nutshell, they can be stated as follows:
\begin{enumerate}
\setcounter{enumi}{-1}
\item The surface gravity is constant on any surface corresponding
to a black hole event horizon.
\item If an amount of material of mass $\delta M$, angular momentum $\delta J$
and charge $\delta Q$ accretes into a black hole, the area of the
event horizon responds according to
\begin{equation}
\frac{\kappa}{8\pi G} \delta A = \Omega\, \delta J + \Phi \,
\delta Q -\delta\, M\, , \label{eq:firstlaw}
\end{equation}
where $\kappa$ is the surface gravity of the horizon, $\Phi$ the
electrostatic potential at the horizon and $\Omega$ the angular
velocity of the horizon.
\item The area, $A$, of the event horizon cannot decrease.
\item It is impossible to reduce $\kappa$ to zero by a finite
sequence of operations.
\end{enumerate}
Although some of these laws have caveats, it was not lost on the
scientists of the day their amazing resemblance to the laws of
thermodynamics. Specifically, the first law of black hole
mechanics and the first law of thermodynamics would be analogous
if one associated entropy with the area and temperature with the
surface gravity. It was J.~Bekenstein who took this analogy most
seriously and today black hole entropy is commonly associated
with his name. (For an interesting survey of black hole thermodynamics see \cite{ref:waldbhthermo}, \cite{ref:padthermo} and references therein.)

Some research at this time started to focus on \emph{quantum}
properties of black holes. A theory of quantum gravity is
notoriously difficult to come by, although today there are some
serious candidate theories. In the late 70s and 80s, in the
absence of a full theory of quantum gravity, scientists started
to study black holes from a \emph{semi-classical} perspective.
That is, the geometry of space-time was treated classically, but
the matter fields propagating on the space-time were treated
quantum mechanically. The fields were usually ``test fields'' in much the
same way as test particles are used. They are quantized on the
background space-time using techniques similar to many of those
employed in quantum field theory in Minkowski space-time. There
are, however, some issues, ambiguities and subtleties in curved
space-time that are not present in the corresponding theory in
flat space-time. Some of this will be discussed in later sections. Excellent expositions of this subject may be found in  \cite{ref:waldbhthermo} and \cite{ref:BandD}.

One very important result that has emerged from semi-classical
studies is that black holes do indeed, as Wheeler suspected,
evaporate. This black hole evaporation was first suggested for
rotating black holes by Zel'dovich and later, in 1974,
Hawking quantitatively discovered that \emph{all} black holes must radiate
\cite{ref:hawkevap}. This was confirmed and extended by D.~Page and
W.~Unruh \cite{ref:pagerad} \cite{ref:unruhrad}.  To this thermal radiation one could associate
a temperature, which for the Schwarzschild hole is given by
\begin{equation}
 T_{Sbh}= \frac{\hbar c^{3}}{8\pi GMk}\:, \label{eq:bhtemp}
\end{equation}
as well as an entropy,
\begin{equation}
 S_{Sbh}\approx \frac{kc^{3}}{4\pi G \hbar} A\:, \label{eq:bhent}
\end{equation}
with $A$ being the area of the black hole's horizon. The
temperature is very small, approximately $10^{-7}$ K for a black
hole of the order of a solar mass. The entropy, on the other
hand, is very large, being of the order of $10^{54}$ for a solar
mass black hole (in J$\cdot$K$^{-1}$). As we will see, theories of quantum gravity may
explain the origin of this large entropy from the gravitational
degrees of freedom associated with the horizon.

Since these results, the arena of semi-classical black hole
research has been a very fruitful one. However, a more
fundamental problem remained; a full theory of quantum gravity was
still (and in many ways still is) elusive.

\section{Classical Black Hole Research}
\subsection{Review of important solutions}
Here we briefly review some of the important classical black hole solutions and their properties. These solutions are amongst the most studied metrics in general relativity theory. A detailed exposition on the mathematical aspects of classical black holes may be found in \cite{ref:chandra} and \cite{ref:fronov}.

\subsubsection{The Kottler-Reissner-Nordstr\"{o}m black hole}
Perhaps the most famous solution to the gravitational field equations is the Schwarzschild metric. With charge and cosmological constant it yields the line element:
\begin{equation}
ds^{2}=-\left(1-\frac{2M}{r}+ \frac{Q^{2}}{r^{2}}-\frac{\Lambda}{3}r^{2}\right)dt^{2} + \frac{dr^{2}}{1-\frac{2M}{r}+\frac{Q^{2}}{r^{2}}-\frac{\Lambda}{3}r^{2}} +r^{2}\,d\theta^{2} +r^{2}\sin^{2}\theta\,d\varphi^{2}\:. \label{eq:krnsol}
\end{equation}
Horizons exist where $g_{tt}=0$.

It is actually not difficult to derive this solution. Consider a general, static, spherically symmetric line element:
\begin{equation}
ds^{2}=-e^{\gamma(r)}\, dt^{2} + e^{\alpha(r)}\, dr^{2} + r^{2}\,
d\theta^{2} + r^{2}\sin^{2}(\theta)\, d\varphi^{2}\:. \label{eq:sphereline}
\end{equation}
Expression (\ref{eq:sphereline}) yields the following, from the field equations (\ref{eq:einsteq}):
\begin{subequations}
\begin{align}
G^{t}_{\;t}=&\frac{e^{-\alpha(r)}}{r^{2}}\left( 1-r\alpha(r)_{,r}
\right)-\frac{1}{r^{2}}= 8\pi T^{t}_{\;t}\label{eq:einstzero} \\
G^{r}_{\; r}=&\frac{e^{-\alpha(r)_{}}}{r^{2}}\left(
1+r\gamma(r)_{,r}
\right)-\frac{1}{r^{2}}= 8\pi T^{r}_{\;r} \label{eq:einstone} \\
G^{\theta}_{\;\theta}\equiv G^{\varphi}_{\;\varphi}=& \frac{e^{-\alpha(r)_{}}}{2} \left(
\gamma(r)_{,r,r} +\frac{1}{2} \left(\gamma(r)_{,r}\right)^{2}
+\frac{1}{r} \left(\gamma(r)_{}-\alpha(r)_{}\right)_{,r} \right.\nonumber \\
&\left.-\frac{1}{2}\alpha(r)_{,r}\gamma(r)_{,r} \right)=8\pi
T^{\theta}_{\;\theta} = 8\pi T^{\varphi}_{\;\varphi}, \label{eq:einsttwo}
\end{align}
\end{subequations}
For simplicity the cosmological term and the charge term are set to zero, however, it is straight-forward to implement these, by including them as part of $T^{\mu}_{\;\nu}$.

Equation (\ref{eq:einstzero}) may be utilized to give the following:
\begin{eqnarray}
e^{-\alpha(r)_{}}&=&\frac{8\pi}{r}\int (r\pme)^{2} \left(T^{t}_{t} (r\pme)\right) dr\pme +1 =:
1-\frac{2 m(r)}{r}\:. \label{eq:alphaeq}
\end{eqnarray}
Since the system of equations is under-determined, two functions may be prescribed. Since the Schwarzschild solution, if considered in the domain $0 \leq r < \infty$, corresponds to the gravitational field of a point mass of mass $M$, we can postulate a stress-energy tensor for a point mass with the following $T^{0}_{\;0}$ component:
\begin{equation}
T^{t}_{\;t}(r)= -\frac{M}{4\pi r^{2}}\, \delta(r)\:.
\end{equation}
The $r$ dependence is motivated by dimensionality arguments and neglecting the factor of $4\pi$ would simply correspond to a rescaling of the mass. There is some arbitrariness on what the other function to be prescribed can be. However, to satisfy junction conditions implied by the above equations supplemented with the conservation law, $T^{r}_{\;r}$ should be continuous, and therefore set to zero. The remaining unknowns may be solved for by straight-forward manipulation of the field equations and conservation law:
\begin{subequations}
\begin{align}
e^{\gam}=&e^{-\alf} e^{h_{0}}, \label{eq:egamma}\\
T^{\theta}_{\;\theta}\equiv T^{\varphi}_{\;\varphi}
:=&\frac{M\,\delta(r)}{16\pi r}\gamma_{,r}\:. \label{eq:pressure}
\end{align}
\end{subequations}
The constant $h_{0}$ can be absorbed into the definition of the time coordinate.

For the case of a charged black hole ($Q\neq0$, $\Lambda=0$) there are two horizons, one at $r=r_{+}=M+\sqrt{M^{2}-Q^{2}}$ and another, inner, horizon at $r=r_{-}=M-\sqrt{M^{2}-Q^{2}}$. In the extremal case, $Q=M$ and the horizons are coincident.

In closing this sub-section we quote the form of the line element in several other well known coordinate systems which historically have shed light on the causal structure (now with $Q=0=\Lambda$). In Painlev\'{e}-G\"{u}llstrand coordinates, the Schwarzschild metric is cast in the form:
\begin{equation}
ds^{2}=-d\tilde{t}^{2} + \left(dr +\sqrt{\frac{2M}{r}}\,d\tilde{t}^{2}\right)^{2} +r^{2}\,d\theta +r^{2}\sin^{2}\theta\,d\varphi^{2}\:. \label{eq:PGcoord}
\end{equation}
These coordinates are regular at the horizon (although $g_{00}$ still vanishes there) and readily display the no-escape property of this surface.
Considering radial light-rays ($ds=0,\, d\theta=d\varphi=0$), the equation of motion yields:
\begin{equation}
\frac{dr}{d\tilde{t}}=\pm 1 -\sqrt{\frac{2M}{r}} \:. \nonumber
\end{equation}
Note that for $r < 2M$ the quantity $\frac{dr}{d\tilde{t}}$ is negative for both solutions, indicating that both ingoing and ``outgoing'' null rays approach $r=0$.

\begin{figure}[h!t]
\begin{center}
\includegraphics[bb=0 25 905 482, clip, scale=0.42, keepaspectratio=true]{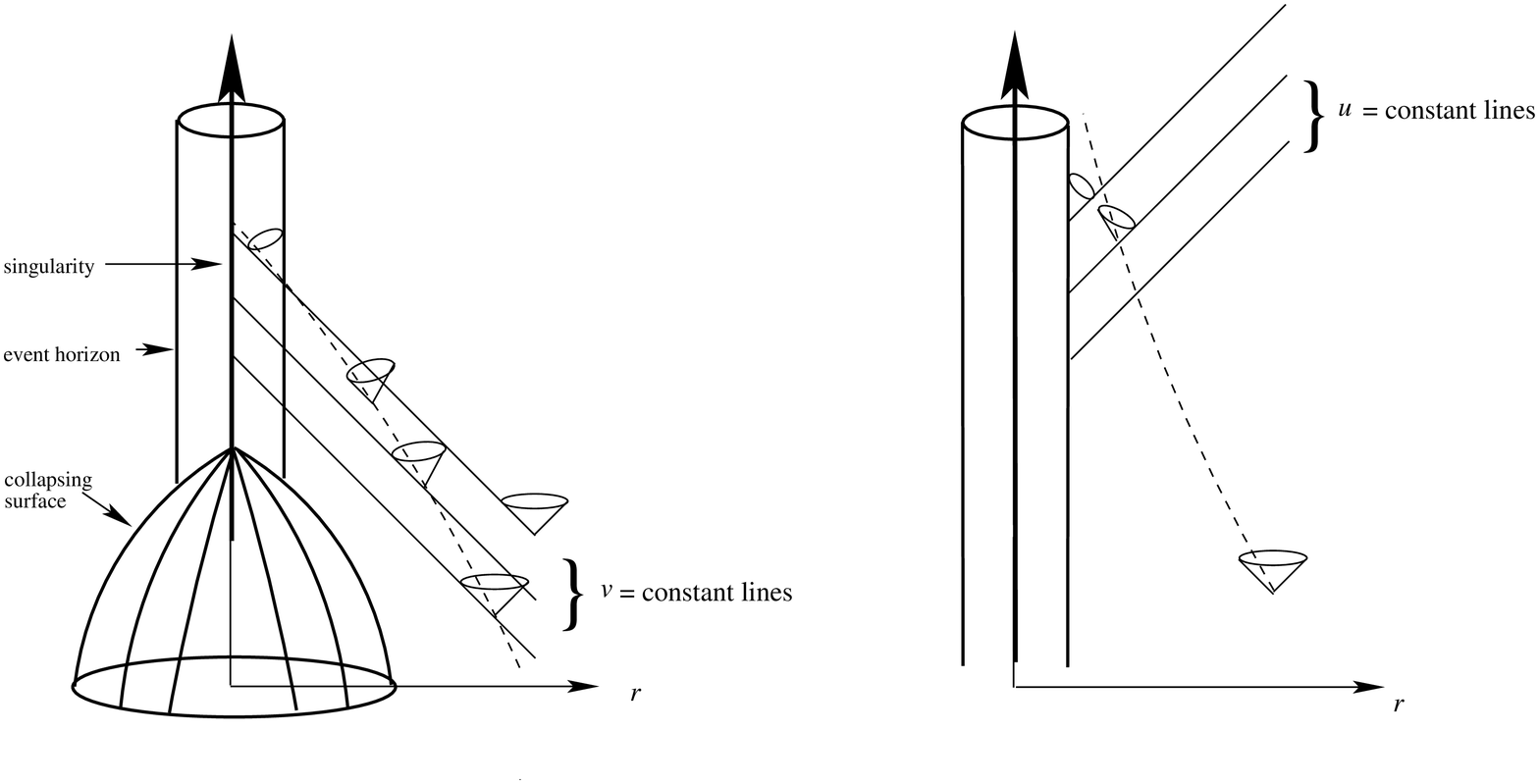}
\caption{{\small The ingoing Eddington-Finkelstein coordinates
(left) and the outgoing Eddington-Finkelstein coordinates (right).
The dashed line represents an in-falling time-like particle.}}
\label{fig:EFcoord}
\end{center}
\end{figure}

The ingoing Eddington-Finkelstein coordinates are also particularly useful in elucidating the ``no escape'' property of the event horizon. In these coordinates, the Schwarzschild metric yields:
\begin{equation}
 ds^{2}=-\left(1-\frac{2M}{r}\right) dv^{2} +2\,dv\,dr+ r^{2}\,d\theta^{2} +r^{2}\sin^{2}\theta\,d\varphi^{2}\:. \label{eq:inEF}
\end{equation}
The nature of these coordinates, along with the causal structure is demonstrated in figure \ref{fig:EFcoord}.
Of course, there also exists the outgoing Eddington-Finkelstein coordinates:
\begin{equation}
ds^{2}=-\left(1-\frac{2M}{r}\right) du^{2} -2\,du\,dr+ r^{2}\,d\theta^{2} +r^{2}\sin^{2}\theta\,d\varphi^{2}\:, \label{eq:outEF} 
\end{equation}
which are also illustrated in figure \ref{fig:EFcoord}.

\begin{figure}[h!t]
\begin{center}
\includegraphics[bb=0 0 495 412, clip, scale=0.45, keepaspectratio=true]{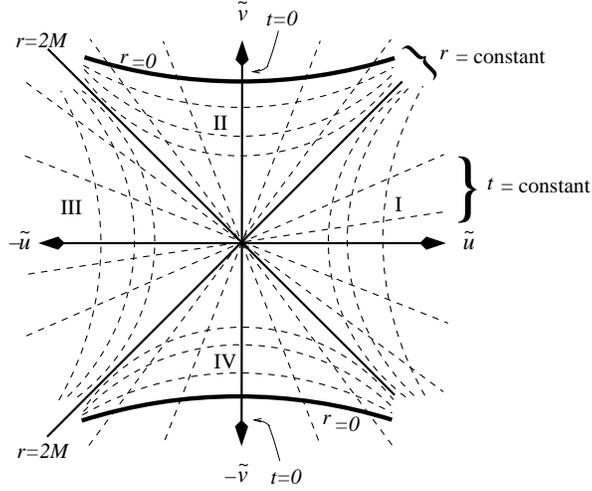}
\caption{{\small The maximally extended Schwarzschild space-time.}}
\label{fig:krusk}
\end{center}
\end{figure}

Finally we present the line element in Kruskal-Szekeres-Synge coordinates, which eliminate the coordinate pathology at $r=2M$ altogether:
\begin{equation}
 ds^{2}=-32 \frac{M^{3}}{r}e^{-r/2M}\,d\tilde{u}d\tilde{v} +r^{2}\,d\theta^{2} +r^{2}\sin^{2}\theta\,d\varphi^{2}\:, \label{eq:kruskline}
\end{equation}
where $r$ now represents the solution to:
\begin{equation}
 \tilde{u}\tilde{v}=\left(1-\frac{r}{2M}\right)e^{r/2M}\:. \nonumber
\end{equation}
The Kruskal diagram (also known as the maximally extended Schwarzschild space-time) is illustrated in figure \ref{fig:krusk}. The two coordinate patches corresponding to $r > 2M$ and $r < 2M$ in metric (\ref{eq:krnsol}) (recall $Q=0$, $\Lambda=0$ in this discussion) are capable of describing regions I and II respectively of the maximally extended Schwarzschild space-time.

Further discussion of horizon-regular coordinate systems may be found in \cite{ref:martpois}. 

\subsubsection{The Kerr-Newman-de Sitter black hole}
This space-time describes a charged rotating ring or, in its asymptotic regime, the space-time outside a rotating charged star. It is the rotational analogue of the Kottler-Reissner-Nordstr\"{o}m black hole. Unlike the previous case, here there is no Birkhoff's theorem \cite{ref:birk} guaranteeing uniqueness of this solution as the exterior of a realistic rotating star.

The solution, without charge or cosmological constant, was discovered by Roy Kerr \cite{ref:kerr} in 1963, although several others had attempted to find such a solution before him. Kerr's original metric yielded the line element:
\begin{align}
 ds^{2}=& -\left(1-\frac{2Mr}{\rho^{2}}\right)\left(du + a\sin^{2}\theta\,d\varphi\right)^{2} \nonumber \\
&+2\left(du + a^{2}\sin^{2}\theta\, d\varphi\right)\left(dr + a\sin^{2}\theta\, d\varphi\right) \nonumber \\
&+\rho^{2} \left(d\theta^{2} +\sin^{2}\theta\, d\varphi^{2}\right), \label{eq:kerrssoln}
\end{align}
where 
\begin{equation}
\rho^{2}:=r^{2}+a^{2}\cos^{2}\theta\:, \label{eq:rhosquared}
\end{equation}
with $a$ the angular momentum per unit mass and $M$ the black hole mass.

With the addition of charge ($Q$), and in the presence of a cosmological constant ($\Lambda$), the line element is expressible in the following formidable form, utilizing the Boyer-Lindquist coordinates:
\begin{eqnarray}
ds^2&=&-\frac{1}{\rho^{2}\Sigma^2}\left(\Delta_{r}-\Delta_{\theta}a^{2}\sin^{2}\theta\right)dt^{2}
+\frac{\rho^{2}}{\Delta_{r}}\,dr^{2}+\frac{\rho^{2}}{\Delta_{\theta}}\,d\theta^{2}\nonumber\\
&&+\frac{1}{\rho^{2}\Sigma^{2}}\left[\Delta_{\theta}\left(r^{2}+a^{2}\right)^{2}-
\Delta_{r}a^{2}\sin^{2}\theta\right]\sin^{2}\theta\, d\varphi^{2}  \nonumber \\
&&-\frac{2a}{\rho^{2}\Sigma^{2}}\left[\Delta_{\theta}{\left(r^{2}+a^{2}\right)}-\Delta_{r}\right]\sin^{2}\theta\,
dt\,d\varphi\:, \label{KNDSmetric}
\end{eqnarray}
with
\begin{subequations}
\begin{align}
\Delta_{r}=&\left(r^{2}+a^{2}\right)\left(1-\frac{\Lambda}{3}r^{2}\right)-2Mr+Q^{2}\:,\label{metric_expb}\\
\Delta_{\theta}=&1+\frac{1}{3}\Lambda
a^2\cos^{2}\theta,\;\;\;\;\;\;\Sigma=1+\frac{1}{3}\Lambda
a^2\:.\label{metric_expc}
\end{align}
\end{subequations}

This metric is quite complicated to work with and we will therefore focus attention on the case $Q=0$ and $\Lambda=0$. It will also be assumed that $a < M$. 

In summary, the $Q=\Lambda=0$ metric possesses the following properties: In the limit $a\rightarrow 0$ this solution goes over to the Schwarzschild solution. The $M=0$ limit yields flat Minkowski space-time in oblate spheroidal coordinates. As well, a true singularity exists at $\rho=0$ as can be seen from the computation of the Kretschmann scalar
\begin{equation}
R_{\alpha\beta\gamma\delta}R^{\alpha\beta\gamma\delta}=\frac{1}{\rho^{12}}48M(r^{2}-a^{2}\cos^{2}\theta)\left[\rho^{4}-16r^{2}a^{2}\cos^{2}\theta\right]\:. \label{eq:kerrkretsch}
\end{equation}
The singularity is located at $r=0,\, \theta=\pi/2$, which in a Cartesian-type coordinate system corresponds to $x^{2}+y^{2}=a^{2},\,z=0$, indicating a ring-like structure to the singularity. Other domains of interest are:\\
i) $r=r_{\pm}:=M\pm\sqrt{M^{2}-a^{2}}$; these are the inner and outer event horizons.  \\
ii) The coordinates $t$ and $\varphi$ are not orthogonal. This implies that a geodesic observer ``moving'' in the time direction must necessarily move in the $\varphi$ direction. That is, the observer is dragged around the black hole in the direction of rotation. This is the famous Lense-Thirring effect \cite{ref:lensthir}.\\
iii) Related to the previous item, $r=r_{\mbox{{\tiny{E}}}_{\pm}}:=M\pm\sqrt{M^{2}-a^{2}\cos^{2}\theta}$ are inner and outer ergosurfaces. A time-like observer cannot resist the dragging effects while in this region. \\
The light cone structure is studied in, for example, \cite{ref:hawkpen}, \cite{ref:chandra} and \cite{ref:schee}.

\begin{figure}[h!t]
\begin{center}
\includegraphics[bb=0 0 886 787, clip, scale=0.4, keepaspectratio=true]{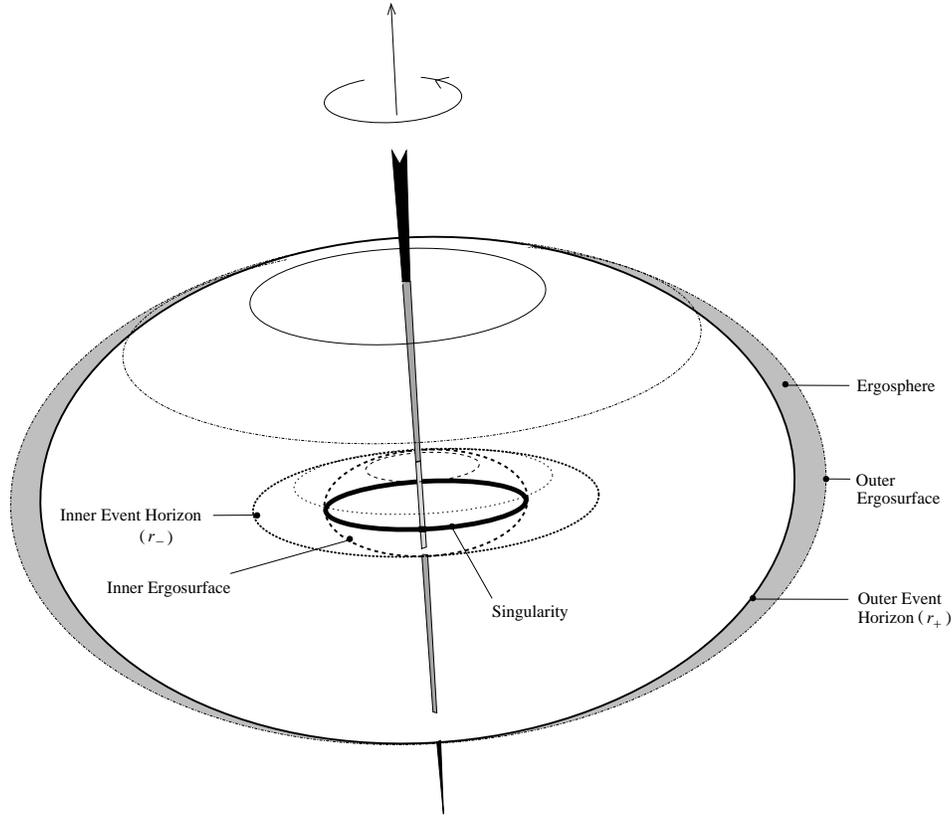}
\caption{{\small Various regions of a Kerr black hole in a
Cartesian-type coordinate system.}} \label{fig:kerrregions}
\end{center}
\end{figure}

Figure \ref{fig:kerrregions} illustrates the relative locations of these domains, in a Cartesian-like set of coordinates. 

There are relatively few treatises on the Kerr geometry, mainly due to the technical difficulties involved in working with the metric. The interested reader is referred to \cite{ref:oneill_kerr}, \cite{ref:scott_vis_wilt}.

\subsubsection{Exotic Solutions}
Here we briefly describe some solutions which are ``exotic'' in some sense. By exotic we mean solutions that are unlikely to exist in nature but are still of great interest on a theoretical basis. If one allows for a \emph{negative} cosmological constant, then solutions exist in general relativity which can describe black holes with \emph{planar}, \emph{cylindrical}, or \emph{higher genus} topology. These are sometimes known as topological black holes. A sufficiently general metric to describe such solutions (here with vanishing charge and angular momentum) is given by:
\begin{equation}
ds^2=-\left(\alpha^2 r^2 -b -\frac{2M}{r}\right)\,dt^{2} + \frac{d\rho^{2}}{\left(\alpha^2 r^2 -b -\frac{2M}{r}\right)} +r^2\left(d\theta^{2} + d\sinh^{2}(\sqrt{b}\,\theta)\,d\varphi^{2}\right)\:, \label{eq:smmetric}
\end{equation}
with $0 < \varphi \leq 2\pi$. Here, $\alpha$ is related to the cosmological constant via $\alpha^{2}=-\Lambda/3$, $M$ is the mass parameter, and $d$ and $b$ are constants that determine the topology of $t,r=$constant surfaces. The cases are as follows:\\
i) $b=-1$, $d=-1$: In this case constant $(t,\,r)$ surfaces are spheres (the Kottler solution). \\
ii) $b=0$, $\underset{b\rightarrow 0}\lim\, d=\frac{1}{b}$: In this case constant $(t,\,\rho)$ surfaces are tori. \\
iii) $b=1$, $d=1$: In this case constant $(t,\,r)$ surfaces are surfaces of constant negative curvature of genus $g > 1$, depending on the identifications chosen. \\
An event horizon exists when $\left(\alpha^2 r^2 -b -\frac{2M}{r}\right)=0$.
Such topologies were studied in detail in \cite{ref:firsttop}. The formation of such black holes from gravitational collapse was studied in \cite{ref:tor1}.

The metric (\ref{eq:smmetric}) does not uniquely describe such topological black holes. Lemos and Zanchin, for example,  have studied a class of black holes which can be cast in the following form \cite{ref:tor2} \cite{ref:tor3}:
\begin{equation}
 ds^{2}=-\left(\alpha^{2}r^{2} - \frac{B_{o}M}{\alpha r}\right)dv^{2} +2\, dv\,dr +r^{2} \left(d\theta^{2} + d\varphi^{2}\right)\:. \label{eq:onenulltopo}
\end{equation}
Here, the values of $B_{0}$ and the identifications make up either toroidal, cylindrical or planar topologies. Specifically: \vspace{0.1cm} \\
i) $\, 0 \leq \theta < 2\pi$, $0 \leq \varphi < 2\pi$, $B_{0}=\frac{2\alpha}{\pi}$ yields the flat torus model. \vspace{0.1cm} \\
ii) $\, -\infty < \theta < \infty$, $0 \leq \varphi < 2\pi$, $B_{0}=4$ yields the cylinder, with $M$ the mass per unit length and the linear axis coordinate, $z$, is related to $\theta$ via $\theta=\alpha z$.\vspace{0.1cm} \\
iii) $\, -\infty < \theta < \infty$, $-\infty < \varphi < \infty$, $B_{0}=\frac{2}{\alpha}$ yields the planar case, with $M$ the mass per unit area. \vspace{0.05cm}\\
This metric differs slightly from the metric (\ref{eq:smmetric}). Gravitational collapse forming such black holes was studied in \cite{ref:tor3}. It can be checked that in all the above cases, the Kretschmann scalar blows up at $r=0$. Another construction of topological black holes may be found in \cite{ref:amine}. Research involving the black holes described in this section will be presented below.

\subsection{Developments in classical black hole research}

\subsubsection{Quasi-normal mode analysis} 
Broadly defined, black hole quasi-normal modes arise from perturbations in some black hole space-time. However, the modes are affected by the emission of gravitational radiation, which generally has a damping effect on the modes, and therefore these modified modes are named quasi-normal \footnote{Here we ignore subtle, but important, mathematical questions regarding the completeness of quasi-normal modes.}. The oscillations can in theory be reconstructed via the analysis of their corresponding gravitational wave emissions. Reversing the argument, the vibrational modes of the black hole are closely linked to the corresponding emitted gravitational wave pattern and they are therefore important in light of gravitational wave astronomy. (In the case of pure gravitational perturbations, the oscillations are by definition the gravitational wave patterns.) It is believed that the gravitational waves due to oscillations produced during black hole formation may be strong enough to detect with current or near future gravitational wave detectors. The wave signature will be unique, the frequencies tending not to depend strongly on the perturbing process and which would yield a direct measurement of a black hole's existence and give information on its mass and angular momentum. Two or more modes may give useful information such as helping discern if general relativity is valid. In particular it may prove or disprove the no-hair theorem of general relativity \cite{ref:bertcardwill}. An excellent expos\'{e} on quasi-normal modes may be found in the thesis by Cardoso \cite{ref:qnmcarthesis} as well as the works \cite{ref:padqnm1} and \cite{ref:padqnm2}. Also, a good reference on non-spherical metric perturbations of the Schwarzschild black-hole spacetime may be found in \cite{ref:rezpert1}.

Perturbations of black holes were originally studied by Regge and Wheeler \cite{ref:regwheel}. For the Schwarzschild black hole one may, for example, perturb the metric in a way appropriate for ``axial'' perturbations:
\begin{equation}
ds^{2}=-\left(1-\frac{2M}{r}\right)\,dt^{2} + \frac{dr^{2}}{1-\frac{2M}{r}} +
r^{2}\,d\theta^{2} +r^{2}\sin^{2}\theta\left[d\varphi -\omega\,dt
-q_{2}\,dr -q_{3}\,d\theta\right]^{2} . \label{pertschw}
\end{equation}

We write
\begin{equation}
\omega(r,\;\theta,\;t)=\tilde{\omega}(r,\;\theta)e^{i\sigma t} , \label{eq:omegapert}
\end{equation}
and similarly for $q_{2}$ and $q_{3}$.

For perturbations of this form, it can be shown that the system
governing the perturbations can be reduced to a single
second-order differential equation, which can be solved by
separating the variables $r$ and $\theta$ (for example, see \cite{ref:chandra}). Briefly, the perturbed field equations give a relation between
$\omega$ and $q_{2}$ and $q_{3}$, allowing the elimination of
$\omega$. The quantity $Q:=(1-2M/r)r^{2}\sin^{3}\theta\left(
\partial_{\theta}q_{2} -\partial_{r}q_{3}\right)$ is written as
$Q=R(r)\Theta(\theta)$ and the resulting radial equation is
\begin{equation}
r^{2}\left(1-\frac{2M}{r}\right)\frac{d}{dr} \left[\frac{1-\frac{2M}{r}}{r^{2}}
\frac{dR}{dr}\right] -\mu^{2}_{l} \frac{\left(1-\frac{2M}{r}\right)R}{r^{2}}
+\sigma^{2} R =0, \label{eq:radeq}
\end{equation}
where $\mu_{l}^{2}$ is the eigenvalue of the angular equation
and may take on the values $\mu_{l}^{2}:=(l+2)(l-1)$ for
$l=2,\;3,\;...\;\;$.

We can make a change of coordinates,
\begin{equation}
r_{*}=r+2M \ln\left(\frac{r}{2M}-1\right) , \nonumber
\end{equation}
so that (\ref{eq:radeq})
reduces to a Schr\"{o}dinger-type equation:
\begin{equation}
\left[\frac{d^{2}}{dr^{2}_{*}} - V_{l}(r)\right]Z_{l}= -\sigma^{2}_{l}Z_{l}\label{eq:schro}
\end{equation}
where
\begin{equation}
Z_{l}(r)=\frac{R_{l}(r_{*})}{r}\;, \nonumber
\end{equation}
and
\begin{equation}
V_{l}(r)=\left(1-\frac{2M}{r}\right) \left[\frac{l(l+1)}{r^{2}} -\frac{6M}{r^{3}}\right].
\label{eq:V}
\end{equation}
The equation (\ref{eq:schro}) is often referred to the Regge-Wheeler equation with (\ref{eq:V}) the Regge-Wheeler potential. The above expression is valid for scalar perturbations, the coefficient of the last term differing for vector and tensor perturbations (see \cite{ref:qnlrr} for details). Later, Zerilli derived a similar equation for polar perturbations with a more complicated potential \cite{ref:zer}:
\begin{equation}
 V_{l}(r)_{pol}=\frac{2(r-2M)}{r^{4}(nr+3M)}\left[n^{2}(n+1)r^{3} +3Mn^{2}r^{2} +9M^{2}nr + 9M^{3}\right], \label{eq:polV}
\end{equation}
with $n:=\frac{1}{2}(l+2)(l-1)$.

One can compute frequencies for functions which possess the asymptotic form \cite{ref:vishvesh}
\begin{equation}
Z_{l}(r) \rightarrow 
\left\{ 
\begin{array}{lll}
e^{i\sigma r_{*}} & \mbox{for} & r_{*}\rightarrow \infty \:, \\
e^{-i\sigma r_{*}} & \mbox{for} & r_{*}\rightarrow -\infty \: .
\end{array}
\right.
\end{equation}
The frequencies are complex and will be denoted as $\sigma=\sigma_{1}+i\sigma_{2}$.
For every value of $l$ there exist a tower of modes, denoted here by the integer $m$. Some of the lowest frequencies for the Schwarzschild black hole are summarized in table 1 which hold for both the axial and polar perturbations.

\setcounter{table}{0}
\begin{table}[h!t]
\caption{{\small A summary of the first four quasi-normal mode
frequencies in units of $M\sigma$ for $l=2,\,3$ and $4$. The
conversion to Hz is given by multiplying the given numbers by $2\pi
\frac{M_{{\mbox{\tiny{$\astrosun$}}}}}{M}5142$Hz. This data is from
\cite{ref:qnlrr} and \cite{ref:qnmgw}.}}
\begin{center}
\begin{tabular}{|l||l|l|l|}
\hline
&$\;\;\;\;\;\;\;\;l=2$ & $\;\;\;\;\;\;\;\;l=3$ & $\;\;\;\;\;\;\;\;l=4$ \\
\hline\hline
& \underline{$\;M\sigma_{1}\;\;\;\;\;\;M\sigma_{2}$}& \underline{$\;M\sigma_{1}\;\;\;\;\;\;M\sigma_{2}$} & \underline{$\;M\sigma_{1}\;\;\;\;\;\;M\sigma_{2}$} \\
$m=0$   & $0.3737\;\;\;0.0890$ & $0.5994\;\;\;0.0927$ & $0.8092\;\;\;0.0942$  \\
$m=1$   & $0.3467\;\;\;0.2739$ & $0.5826\;\;\;0.2813$ & $0.7966\;\;\;0.2844$  \\
$m=2$   & $0.3011\;\;\;0.4783$ & $0.5516\;\;\;0.4791$ & $0.7727\;\;\;0.4799$  \\
$m=3$   & $0.2515\;\;\;0.7051$ & $0.5120\;\;\;0.6903$ & $0.7398\;\;\;0.6839$  \\
\hline
\end{tabular}
\end{center}
\end{table}

\begin{figure}[h!t]
\begin{center}
\includegraphics[bb=0 0 294 391, clip, angle=-90, scale=0.55, keepaspectratio=true]{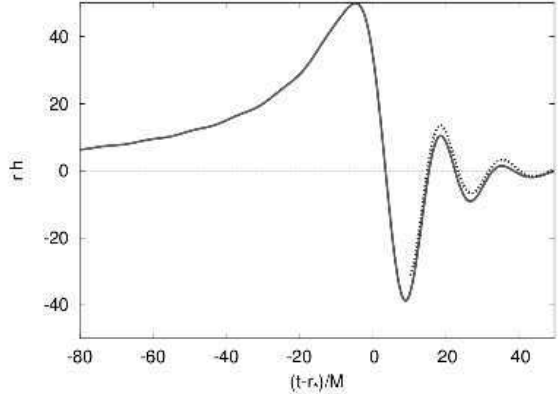}
\caption{{\small A computation of the gravitational wave amplitude of a Schwarzschild black hole perturbed by an infalling particle (solid line) along with an analytic fit (dashed line) utilizing the first two modes corresponding to $l=2$. Figure courtesy of V. Ferrari and L. Gualtieri, Roma. (From \cite{ref:qnmgw}.) (figure quality reduced for arXiv file size)}} \label{fig:qnm}
\end{center}
\end{figure}

One item of particular interest in modern quasi-normal mode research is that of radiative tails. The tail refers to the non-trivial fall-off properties (in time) of the perturbation. This phenomenon is sometimes known as black hole ringing, due to the oscillatory behavior of the tail. An example of such a tail is given in figure \ref{fig:qnm}, for an infalling particle perturbing a Schwarzschild black hole. The vertical axis represents the gravitational wave amplitude. The dashed line represents an analytical fit using a linear combination of the first two $l=2$ modes.

The Kerr black hole is much more difficult to analyze. Some work has been performed in \cite{ref:qnmkerrdet} - \cite{ref:qnmkerrono}. Studies find that as the angular momentum increases, $\sigma_{1}$ is bounded, whereas $\sigma_{2}$ is not. However, as $a$ approaches the value of $M$, $\sigma_{2}$ tends to zero and the oscillations would therefore continue without damping \cite{ref:qnmkerrdet}. These modes may not be realized though as some studies indicate that the amplitudes of these modes also tend to zero \cite{ref:qnmkerrmash}. 

Another interesting phenomenon associated with the Kerr black hole is that of superradiance. For massless vector and tensor perturbations due to scattering off of the Kerr potential, the reflection coefficient can exceed unity if the incoming wave possesses a frequency below a certain critical value. This is believed to be due to an interplay between particle creation in black holes and the Penrose energy extraction process \cite{ref:qnmkerrstar} \cite{ref:qnmkerrpress}.

The understanding of quasi-normal modes from black holes is becoming a very important issue in black hole physics due to the possibility of detection with gravitational wave detectors. There are several astrophysical processes that can give rise to quasi-normal modes. For example, an infalling particle could provide such a perturbation. The perturbation due to infalling extended bodies has also been studied \cite{ref:qnmext}. In this case, the effect is smaller than in the point particle case due to interference effects. In the case of fluid material orbiting the black hole, modes are only significantly excited when $r < 4M$, and is therefore important in the case of unstable orbits. More realistic processes involve large scale computations and make up an important area of study in modern black hole research. The collapse of a neutron star core has been studied \cite{ref:qnmns} as have the modes caused by thick accretion disks \cite{ref:qnmad1} and \cite{ref:qnmad2}. We will discuss the case of black hole mergers separately below.

\subsubsection{Critical behavior}
\begin{figure}[h!t]
\begin{center}
\includegraphics[bb=0 0 755 204, clip, scale=0.36, keepaspectratio=true]{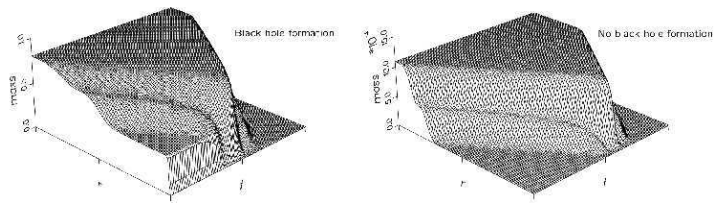}
\caption{{\small Evolution of scalar field data undergoing gravitational collapse. In the left diagram a black hole forms at late time whereas in the right diagram the field disperses. Figures courtesy of M. W. Choptuik, Yukawa International Seminar. (figure quality reduced for arXiv file size)}} \label{fig:critform}
\end{center}
\end{figure}

From the point of view of classical black hole physics there are two possible outcomes that result from the evolution of regular Cauchy data: Either a black hole forms or it does not. In this paradigm, an interesting question to ask is what happens in the regime that straddles this bifurcation? This question was originally studied by Choptuik utilizing a spherically symmetric massless scalar field minimally coupled to gravity \cite{ref:choporig}. This choice of matter field is convenient as one does not need to worry about possible formation of field condensates (stars) in this model. Complete field dispersion or black hole formation are the only possible outcomes. Samples of the two outcomes are displayed in figure \ref{fig:critform}.

The problem was tackled as follows: The initial data contained one tunable parameter, usually denoted as $p$. By changing this parameter the numerical evolution would either form a black hole or not. Ideally, the problem is set up so that there is a simple relationship between the size of this parameter's numerical value and the ``strength'' of the gravitational interaction as there is no way a priori to know if the evolution will result in an event horizon. Therefore many evolutions need to be run, in each trial changing the value of the parameter, before the critical point is reached. 

For the minimally coupled massless scalar, $\phi$, the field equations yield:
\begin{equation}
R_{\mu\nu}-\frac{1}{2}R\,g_{\mu\nu}=8\pi\left(\phi_{,\mu}\phi_{,\nu} -\frac{1}{2} \phi_{,\alpha}\phi^{;\alpha}\, g_{\mu\nu}\right)\:, \label{eq:scalareinst}
\end{equation}
whereas the conservation law gives rise to the wave equation:
\begin{equation}
\phi_{,\alpha}^{\;\,;\alpha}=0\:. \label{eq:waveeq}
\end{equation}

With a metric admitting the line element as in (\ref{eq:sphereline}), except that now $e^{\gamma(r)} \rightarrow e^{\gamma(r,\,t)}=:b^{2}$ and $e^{\alpha(r)} \rightarrow e^{\alpha(r,\,t)}=:a^{2}$, the field equations go over to:
\begin{subequations}
\begin{align}
\frac{a_{,r}}{a} +\frac{a^{2}-1}{2r} - 2\pi r\left[\frac{a^{2}}{b^{2}}(\phi_{,t})^{2} + \phi_{,r}\right]=&0\:, \label{eq:scaleinst1} \\
\frac{b_{,r}}{b}-\frac{a_{,r}}{a}-\frac{a^{2}-1}{r}=&0\:, \label{eq:scaleinst2} \\
\frac{a_{,t}}{b}-4\pi r \phi_{,r} \frac{a}{b} =&0\:. \label{eq:scaleinst3}
\end{align}
\end{subequations}
The wave equation, after some mild manipulation yields the condition:
\begin{equation}
\left(\frac{a}{b}\phi_{,t}\right)_{,t}=\frac{1}{r^{2}}\left(r^{2}\frac{b}{a}\phi_{,r}\right)_{,r}\:. \label{eq:scalarwave}
\end{equation}
Dirichlet and Neumann boundary conditions were used for the metric and the scalar field. At $r=0$, the condition of space-time regularity was imposed. As well, the initial ($t=0$) scalar field profile is also fully specified along with an initial velocity.

Choptuik found some unexpected and very interesting behavior when studying the above system of equations. In the cases that were barely super-critical (i.e. cases where the parameter was tuned to values just above the black hole formation limit), Choptuik found an unexpected relationship. He found that the masses of the ``mini'' black holes formed followed a scaling law of the form
\begin{equation}
M=C(p-p_{c})^{\eta}\:. \label{eq:critmass}
\end{equation}
Here, $p_{c}$ is the critical value of the adjustable parameter. That is the value (within numerical precision, which was $10^{-15}$ for the original study) $p$ attains when it is exactly at the bifurcation point. $C$ is a constant that depends on the particular initial data profile chosen and, perhaps most interestingly, the exponent $\eta$ possesses a value that is \emph{universal} for all initial scalar field data\footnote{This exponent is often labeled as $\gamma$ in the literature. However, in this manuscript $\gamma$ plays multiple roles and therefore $\eta$ is used to avoid confusion.}. For the minimally coupled, massless scalar field, Choptuik found that $\eta \approx 0.374\:$. No matter what the initial data profile, the masses of small black holes followed the law (\ref{eq:critmass}) with this numerical value of the exponent. 

Since Choptuik's original work, various authors have extended the domain of study to other coordinate systems \cite{ref:nullcoord}, different matter models (\cite{ref:matcriti} - \cite{ref:matcritf}) and different numbers of dimensions (\cite{ref:matcritperf}, \cite{ref:dimcriti} - \cite{ref:dimcritf}). Also, work in non-spherical symmetry has been done as well as other couplings and some analytic analysis of the problem (\cite{ref:othercriti} - \cite{ref:othercritf}). Some of the possibilities and results are summarized in table 2. Further, the collapse can be classified as type-I, where there is a minimum finite black hole size, or as type-II, where it is possible to form arbitrarily small mass black holes. The phenomenon discussed here applies to the type-II case.

\begin{table}[h!t]
\caption{{\small A summary of some fields studied and their critical
exponents.}}
\begin{center}
\begin{tabular}{ll}
\hline
Field & $\;\;\;\;\;$ $\eta$ \\
\hline\hline
Massless real scalar field   & $\;\;\;\;\;$ 0.37  \\
Massless complex scalar field   & $\;\;\;\;\;$ 0.39   \\
Massless complex scalar field (angular momentum) & $\;\;\;\;\;$ 0.11   \\
Massive real scalar field (small mass)  & $\;\;\;\;\;$ 0.37   \\
Radiation   & $\;\;\;\;\;$ 0.36   \\
SU(2)   & $\;\;\;\;\;$ 0.20   \\
Gravitational waves   & $\;\;\;\;\;$ 0.36   \\
\hline
\end{tabular}
\end{center}
\end{table}

The behavior discussed in this section is reminiscent of certain phase transitions in statistical mechanics. One example is the liquid-gas phase transition. Near the critical temperature, $T_{c}$, a substance on its boiling curve will possess a discontinuity in the density of the two phases of the form
\begin{equation}
 \rho_{l}-\rho_{g} \propto \left(T_{c}-T\right)^{\eta} \:, \label{eq:liqgas}
\end{equation}
where $\rho_{g}$ and $\rho_{l}$ are the densities in the gas and liquid phases respectively. Ferromagnetic materials obey a similar law near the Curie temperature, $T_{c}$,
\begin{equation}
 m\propto \left(T_{c}-T\right)^{\eta}\:,\label{eq:magnet}
\end{equation}
where $m$ represents the magnitude of the magnetization vector.

There is another interesting phenomenon associated with the threshold of black hole formation, namely that of self-similarity of the space-time. As this phenomenon is mainly of interest on the side of the transition where black holes do not occur (i.e. slightly sub-critical) it is beyond the scope of this review and is omitted due to article size considerations. As well, we have only scratched the surface in citing the large number of works in this field. The interested reader is referred to the thorough reviews in \cite{ref:gundlrr} and \cite{ref:gundphysrep} and references therein.

In closing this section we quote the words of C. Gundlach \cite{ref:gundlrr} on this topic ``Critical phenomena are arguably the most important contribution from numerical
relativity to new knowledge in general relativity to date.''

\subsubsection{Black hole mergers}
The two-body problem in general relativity is extremely difficult to analyze, mainly due to the non-linearity of the field equations. Any, even somewhat realistic, system needs to be evolved numerically in the strong-field regime and even with modern computational technology it is still a taxing problem. Black holes, being ``simple'' objects, and having the possibility of producing a strong enough gravitational wave signal to detect are natural objects to consider in the two body problem. This system provides an excellent arena to study the strong-field effects in G. R. without the complications introduced by material (i.e. non-gravitational) effects. Since the two black hole system is unstable, it is expected that at late times the solution will approach that of a Kerr black hole. The case of a black hole-neutron star merger in full general relativity has been recently analyzed in \cite{ref:shibtan}.

Binary black hole systems were first numerically modeled in 1964 by Hahn and Lindquist \cite{ref:hahnlind} on a $51\hspace{-0.1cm}\times\hspace{-0.1cm}151$ mesh utilizing axial symmetry in a head-on collision scenario. Within 4 hours the evolution had proceeded 50 time steps in their simulation. A decade later Smarr \cite{ref:smarrbin} and Eppley \cite{ref:eppbin} also constructed simulations of head-on collisions in the hope of studying the gravitational wave emissions. With the announcement in 1990 of the LIGO gravitational wave observatories to be constructed, gravitational wave simulations and the two body problem were taken up in force. In such studies the numerical methods almost invariably involve some form of $3+1$ split or null variants of it.

The black hole merger is usually separated into 3 regimes: The inspiral stage, the merger stage and, finally, the ringdown stage (see \cite{ref:pret} for full details, which includes an earlier fourth stage, the Newtonian phase). 

In the inspiral stage, gravitational wave emission has the greatest effects on the dynamics. It is expected that in the case of small to medium mass ratios, large eccentricities in the orbit will decay, yielding a roughly circular orbit by the end of this phase \cite{ref:binpeters1}, \cite{ref:binpeters2}. Extreme mass ratios are expected to possess high eccentricities \cite{ref:eccent1}, \cite{ref:eccent2}. This phase is often well modeled by post-Newtonian and higher-order methods. If the mass ratio is extreme, the smaller partner may be viewed as a test particle in the background space-time of the larger partner. Methods exist to calculate gravitational wave emission in this ``test-particle'' case (see, for example \cite{ref:bintest} and references therein).

An interesting point is that the orbit of the small companion in an extreme mass-ratio scenario will not generally lie in a plane, due to frame-dragging effects. Therefore, the geometry of the surrounding space-time can be well probed by the small companion and this information would be transmitted by the gravitational wave emission which is believed to lie within the future LISA detector's bandwidth.

The merger stage is extremely complicated and requires full numerical investigations. This is a very short lasting phase, perhaps lasting two cycles of the emitted gravitational waves. The luminosity of gravitational waves emitted at this stage may approach the order of $10^{52}$ J/s and the frequency of the gravitational wave approaches that of the dominating quasi-normal mode of the resulting black hole. An enormous amount of energy is liberated in this phase which totals in the neighborhood of three percent of the rest mass energy of the system \cite{ref:pret}.

Finally, the ringdown phase refers to the settling down of the single black hole formed to a Kerr black hole. Quasi-normal mode analysis is quite important in this stage as one has a perturbed black hole space-time radiating away energy via gravitational wave emission. The ringdown is dominated by the following frequencies \cite{ref:bertcardwill}, \cite{ref:pret}, \cite{ref:echev}:
\begin{subequations}
 \begin{align}
\frac{\sigma_{1}}{2\pi}\approx &(32 \mbox{ kHz}) \frac{\sunmass}{M} \left[1- 0.63 \left(1-\frac{J}{M^{2}}\right)^{0.3}\right]\:, \\
\sigma_{2}^{-1} \approx & (20 \mu\mbox{s}) \frac{M}{\sunmass} \frac{1}{\left(1-\frac{J}{M^{2}}\right)^{0.45} \left[1-0.63\left(1-\frac{J}{M^{2}}\right)^{0.3}\right]}\:,
 \end{align}
\end{subequations}
with $J$ the angular momentum of the final resulting black hole. The equivalent of one or two percent of the rest mass energy is radiated during ringdown. At late time the emission is dominated by the radiative power-law tail.

As discussed above, much work in this area is numerical, due to the complications presented by the Einstein field equations. However, there exist constraints in the system of equations which must hold throughout any evolution scheme for it to be valid and this complicates the numerical evolution. Mathematically, these constraints arise as follows: Consider a metric with elements $g_{\mu\nu}$ and define a unit normal, $n_{\mu}\:$, to a class of time slices. We also define $\tilde{g}_{\mu\nu}:= g_{\mu\nu}-n_{\mu}n_{\nu}$. On a space-like hypersurface the following (Gauss-Codazzi) relations hold:
\begin{subequations}
 \begin{align}
  \tilde{R}^{\mu}_{\;\nu\rho\sigma}=& \h^{\mu}_{\;\lambda} R^{\lambda}_{\;\alpha\beta\gamma}\,\h^{\alpha}_{\;\nu} \,\h^{\beta}_{\;\rho} \,\h^{\gamma}_{\;\sigma} - K^{\mu}_{\;\rho}K_{\nu\sigma} + K^{\mu}_{\;\sigma} K_{\nu\rho}\:, \label{eq:GC1}\\ 
D_{\mu}K_{\nu\sigma}- D_{\nu}K_{\mu\sigma}=& -n_{\lambda} R^{\lambda}_{\;\alpha\beta\gamma}\, \h^{\alpha}_{\;\sigma} \,\h^{\beta}_{\;\mu} \,\h^{\gamma}_{\;\nu}\:.  \label{eq:GC2}
 \end{align}
\end{subequations}
Here $\tilde{R}^{\mu}_{\;\nu\rho\sigma}$ is the curvature tensor constructed with $\h$, $D$ is built with the connection of $\h$,  and $K_{\mu\nu}$ is the extrinsic curvature of the hyper-surface. It should be noted that these relations hold regardless of the field equations. By pulling back (\ref{eq:GC1}) and (\ref{eq:GC2}) to the hyper-surface, and utilizing the field equations, one obtains the constraints for the Cauchy problem in general relativity. Therefore, the constraint equations enforce solutions to be the allowable data sub-sets permitted within general relativity theory.

It is a major focus of modern numerical research to find a scheme which ensures that the constraints are not seriously violated in the subsequent evolution. As the continuum ADM form of the field equations are weakly hyperbolic, this is no trivial task and generally some clever method needs to be devised such as consistent modifications to the equations of motion. An evolution program often ``crashes'' or becomes ``unstable'' when constraint violating modes grow to a size that is considered unacceptably larger than discretization/truncation errors. A breakthrough in the long-evolution stability problem occurred recently in 2005-06 with the advent of two methods that yield particularly stable evolutions (at least relatively). These are the generalized harmonic coordinates with constraint damping (GHCCD) method \cite{ref:pretprl} \cite{ref:pret06} and an improved variant of the Baumgarte-Shapiro-Shibata-Nakamura method with moving punctures (BSSN) \cite{ref:bssn1}, \cite{ref:bssn2}, \cite{ref:bssn3}, \cite{ref:bssn4}. Briefly, in the GHCCD method one adds to the field equations a function of the constraints which is designed to minimize or dampen the constraint violation. A recent proposal for such a counter-term in the case of the harmonic coordinates was devised in \cite{ref:gundconst}. In the BSSN scenario one defines a conformal metric,
\begin{equation}
 \bar{g}_{ij}:=e^{-\Phi}\h_{ij},\;\;\;e^{\Phi}=\h^{1/3}\:,
\end{equation}
with $\h$ the spatial three-metric in the ADM decomposition, as well as a conformal trace-free extrinsic curvature quantity
\begin{equation}
 \bar{A}_{ij}:=e^{-\Phi} \left[K_{ij}-\frac{1}{3}\h_{ij}K\right],
\end{equation}
with $K$ the trace of the extrinsic curvature. The conformal connection is given by
\begin{equation}
 \bar{\Gamma}^{i}:=\bar{g}^{jk}\bar{\Gamma}^{i}_{\;jk}\:. \label{eq:barconn}
\end{equation}

In the BSSN scheme $\Phi$, $\bar{A}_{ij}$, $\bar{\Gamma}^{i}$ and the lapse and shift are considered the basic variables of the evolution. The reason this scheme is utilized is that the long ranged degrees of freedom can easily be isolated from the non-radiative ones. Another reason is that the constraints can easily be substituted into some of the evolution equations thus implementing some of the constraints at the dynamic level. Also, with appropriate implementation of gauge, the evolution equations are hyperbolic, which is related to the fact that the connection (\ref{eq:barconn}) is treated as an independent quantity. The ``moving punctures'' refer to the fact that the black hole singularities are represented by punctures which move within the grid although there is no evolution at the puncture itself.

An excellent review of the progess in black hole mergers may be found in Pretorius' review \cite{ref:pret}. We summarize here some recent results.

For the scenario involving two equal mass black holes with minimal spin and eccentricities the amount of energy released in the last stages before a Kerr black hole remnant is approximately 3.5\% of the system's total energy. The resulting Kerr black hole possesses a spin parameter of $a\approx 0.69\:$ (\cite{ref:pret}, \cite{ref:bssn3}, \cite{ref:bssn4}, \cite{ref:bakcent} and references therein). After the ``collision'' the gravitational waveform is dominated by the fundamental of the quadrupole moment of the quasi-normal mode of the final black hole. Also, when the flux is near its maximum, and subsequently, the waveform may be closely represented as a sum of quasi-normal modes, which is surprising as this is expected to be a highly non-linear regime.

If one removes the restriction of equal mass, but maintains minimal eccentricity and spin there is a decrease in the total energy emitted as well as the final spin of the resulting black hole. Also, although the quadrupole is still dominant, higher modes become non-negligible due to the reduction in the symmetry of the problem. This symmetry reduction is also responsible for an asymmetric gravitational radiation beaming. This delivers a recoil to the produced black hole in the orbital plane as there is net momentum carried away by the radiation. The maximum velocity imparted due to this effect seems to be approximately 175 Km/s when the mass ratio is 3:1 \cite{ref:pret}.

For the case of equal mass, nominal eccentricity but non-minimal spin, a new degree of freedom is introduced in this case as the individual black hole spins can be aligned in various directions compared to the orbital angular momentum. If the net spin angular momentum has a component parallel to the orbital angular momentum then more energy will be emitted than in the corresponding zero-spin case. If there is a component anti-parallel, less energy is emitted. Some particular studies indicate that a pair of holes, each with $a\approx 0.76$, radiated approximately $7\%$  of their rest mass energy when the spins were aligned with the orbital angular momentum. In the case where the spins were anti-aligned, only approximately $2\%$ of the rest mass energy was radiated. The final black hole spins were approximately $0.89$ and $0.44$ respectively \cite{ref:spinrad1}, \cite{ref:spinrad2}.

The next case presented is for two equal mass black holes with minimal spin but large eccentricity. This is the case studied in the first complete merger simulation by Pretorius in \cite{ref:pretprl}. In this study two localized scalar field profiles were initially employed which collapsed to form black holes. The outcome of the collapse essentially yields a two black hole vacuum. The two field profiles were given equal magnitude but opposite direction boosts, with zero boost corresponding to a head-on collision. The final result, merger or separation, depends on the value of the single parameter, $k$, measuring the strength of the boost. An interesting result was found. Near the threshold value of the boost parameter, for a given class of initial profiles, the number of orbits, $n$ scale as 
\begin{equation}
 e^{n} \propto \left|k-k_{c}\right|^{-\beta}\:, \label{eq:mergescal}
\end{equation}
where $k_{c}$ is the threshold value of this parameter. The exponent $\beta$ was found to possess a value of approximately $0.34$. As it was noted in \cite{ref:pret}, this behavior is similar to that of test particles in equatorial orbit around a Kerr black hole. Those test particles which are near the capture threshold approach unstable circular orbits of the Kerr space-time and possess a scaling behavior similar to (\ref{eq:mergescal}).

Recently, there have been studies providing simple formulas for the calculation of 
the final spin in a binary merger \cite{ref:rezmerg1}, \cite{ref:rezmerg2}.

\section{Semi-Classical Black Hole Research}
\subsection{Review of semi-classical theory}
Before discussing semi-classical relativity, we shall need a
result from flat space-time for future use. 
Let us begin by studying the real massless scalar field in Minkowski space-time. Such a field obeys the wave equation:
\begin{equation}
 \phi_{,\mu}^{\;\;;\mu}=0. \label{eq:flatwaveeq}
\end{equation}
The field can be decomposed into Fourier components of positive and negative frequency
\begin{equation}
 \phi=\sum_{n=0}^{\infty} \left[a_{n} f_{n}(\mathbf{x}) e^{-i\omega_{n} t} + a^{\dagger}_{n}f^{*}_{n}(\mathbf{x}) e^{i\omega_{n} t}\right] \:. \label{eq:flatfdecomp}
\end{equation}
As is well known, the number of particle excitations associated with a particular state may be deduced from the number operator:
\begin{equation}
 \hat{N}=\sum_{n}a^{\dagger}_{n}a_{n}. \label{eq:flatnumber}
\end{equation}
Under proper Lorentz transformations we have (dropping subscripts on $\omega$)
\begin{equation}
\sum_{n=0}^{\infty} \left[a_{n} f_{n}(\mathbf{x}) e^{-i\omega t} + a^{\dagger}_{n}f^{*}_{n}(\mathbf{x}) e^{i\omega t}\right] = \sum_{m=0}^{\infty} \left[a^{\prime}_{m} f^{\prime}_{m}(\mathbf{x^{\prime}}) e^{-i\omega^{\prime} t^{\prime}} + a^{\dagger\prime}_{m}f^{*\prime}_{m}(\mathbf{x^{\prime}}) e^{i\omega^{\prime} t^{\prime}}\right]\:, \label{eq:ltwaves}
\end{equation}
where the prime denotes quantities calculated in the boosted frame. The relation between the primed and unprimed modes is
\begin{align}
f^{\prime}_{m}=&\sum_{n}\left[\alpha_{mn}f_{n}+\beta_{mn}f_{n}^{*}\right], \nonumber \\
f_{n}=&\sum_{m}\left[\alpha^{*}_{nm}f^{\prime}_{m} -\beta_{nm}f^{\prime *}_{m}\right]\:, \label{eq:wavetransf}
\end{align}
where $\alpha$ and $\beta$ are the Bogoliubov coefficients of the transformation.

The unprimed and primed vacua are defined via:
\begin{equation}
 a_{n}\left|0\right>=0,\;\; a_{m}^{\prime}\left|0^{\prime}\right>=0\:. \label{eq:ltvacs}
\end{equation}
In particular, the unprimed and primed creation and annihilation operators are related as
\begin{equation}
 a_{n}=\sum_{m} \left[\alpha_{nm}a^{\prime}_{m} + \beta^{*}_{nm}a^{\prime\dagger}_{m}\right]\: \label{eq:bogtransf}
\end{equation}
under Lorentz transformations, as can be seen by inserting (\ref{eq:wavetransf}) into (\ref{eq:ltwaves}). Note that if the $\beta$ coefficients are not zero, $\left|0^{\prime}\right> \neq \left|0\right>$ and the two vacua will {\emph not} be the same. That is, one observer's zero particle state will not be a zero particle state for a Lorentz transformed observer. Another way to view this is that the number operator in the primed frame does not agree with the number operator in the unprimed frame. In Minkowski space-time, proper Lorentz transformations respect the condition $\beta_{nm}=0$ and therefore uniformly boosted observers will agree on particle content. This is not true for the case of observers which are accelerating, even in flat space-time. This leads to an interesting effect known as Unruh radiation. Discussion of this is beyond the scope of this manuscript.

The semi-classical approach is based on the premise that the gravitational field remains classical, but the matter content is quantized. The most straight-forward way to incorporate this into Einstein's theory is to write Einstein's equations as
\begin{equation}
 R_{\mu\nu} - \frac{1}{2}R\,g_{\mu\nu}= 8\pi \left<\psi\right|T_{\mu\nu}(\phi_{i})\left|\psi\right>=: 8\pi\quantt \:. \label{eq:semiclasseinst}
\end{equation}
That is, the expectation value of a stress-energy tensor operator that depends on quantized fields, $\phi_{i}$, is utilized as a source term in the gravitational field equations. Of course, one may still add a purely classical piece to the right-hand-side of the equations, $T_{\mu\nu}$\,, so that $\left<\psi\right|T_{\mu\nu}(\phi_{i})\left|\psi\right>$ yields quantum corrections to the classical theory. The simplicity of equation (\ref{eq:semiclasseinst}) is deceptive. The matter fields themselves are, of course, quantized on the curved space-time, generally leading to a complicated dependence on $g_{\mu\nu}$ on the right-hand-side. There is also the issue of what state the fields should be in. As well, the quantity $\quantt$ will diverge and therefore a regularization of the effective action leading to $\quantt$ needs to be performed.

There are several states of particular importance in semi-classical black hole physics. These include, but are certainly not limited to, the Hartle-Hawking vacuum, the Unruh vacuum, and the Boulware vacuum. The Hartle-Hawking vacuum corresponds to a black hole in thermal equilibrium with a bath of thermal radiation, the Unruh vacuum corresponds to a state with a particle flux at future infinity, and the Boulware vacuum to a state where particles do not traverse to infinity, as measured from near the black hole. 

In the case of a scalar field, the regularization leads to a renormalized value of the cosmological constant and a renormalized value of the Newtonian gravitational constant. As well, the divergence in the effective action also leads to a term possessing fourth-order derivatives of the metric and terms quadratic in the curvature as the divergences possess the form:
\begin{equation}
 \quantt_{div}\propto  \frac{1}{\sigma}\left[A g_{\mu\nu} + B G_{\mu\nu}\right] +\left({_{(1)}C}\, {_{(1)}H}_{\mu\nu}+{_{(2)}C}\,{_{(2)}H}_{\mu\nu}\right) \ln \sigma \:.
\end{equation}
Here ${_{(1)}H_{\mu\nu}}$ and ${_{(2)}H_{\mu\nu}}$ are given by:
\begin{subequations}
\begin{align}
 {_{(1)}H_{\mu\nu}}:=& 2 R_{;\mu;\nu}-2 g_{\mu\nu}R^{;\rho}_{\;;\rho} +\frac{1}{2} g_{\mu\nu} R^{2} -2R R_{\mu\nu}\:, \\
{_{(2)}H_{\mu\nu}} :=& 2 R_{\mu\;\;;\nu;\alpha}^{\;\:\alpha} - R_{\mu\nu\;\;;\rho}^{\;\;\;\:;\rho} -\frac{1}{2} g_{\mu\nu} R^{;\rho}_{\;;\rho} -2R_{\mu}^{\;\:\alpha}R_{\alpha\nu} +\frac{1}{2} g_{\mu\nu} R^{\alpha\beta}R_{\alpha\beta}\:,
\end{align}
\end{subequations}
and $\sigma$ is Synge's world function \cite{ref:synge}, which is equal to one-half the square of the geodesic distance between two nearby points. The limit $\sigma \rightarrow 0$ needs to be taken and the logarithmic divergence may be removed by renormalization of the constants ${_{(1)}C}$ and ${_{(2)}C}$. Although these terms are due to the metric dependence of $\quantt$ itself, they are often written on the left-hand-side of the field equations:
\begin{equation}
 G_{\mu\nu(ren)}+\Lambda_{(ren)} g_{\mu\nu} + {_{(1)}C}_{(ren)}\,{_{(1)}H_{\mu\nu}} + {_{(2)}C}_{(ren)}\,{_{(2)}H_{\mu\nu}} = 8\pi \left[T_{\mu\nu}+\quantt_{(ren)}\right]\:, \label{eq:reneinst}
\end{equation}
where the subscript $(ren)$ denotes renormalized values. Equations (\ref{eq:reneinst}) are sometimes known as the semi-classical Einstein field equations.

An approximation often employed in semi-classical research is a perturbative approach. That is, the metric is written as:
\begin{equation}
 g_{\mu\nu}= {_{(0)}g}_{\mu\nu}+ \epsilon h_{\mu\nu}\:,
\end{equation}
where ${_{(0)}{g}_{\mu\nu}}$ is a classical ``background'' metric and the small constant parameter $\epsilon$ vanishes in the limit $\hbar \rightarrow 0$ so that $h_{\mu\nu}$ represents first-order quantum corrections to the classical metric. One customarily limits calculations to order linear in $\epsilon$. The Einstein equations then yield an Einstein tensor of the form:
\begin{equation}
 G_{\mu\nu}= {_{(0)}G_{\mu\nu}}+ \epsilon \, {_{(\hbar)}G_{\mu\nu}}\:,
\end{equation}
with ${_{(\hbar)}G_{\mu\nu}}$ being due to the quantum part of the metric and ${_{(0)}G_{\mu\nu}}$ satisfies the usual Einstein equation for the classical stress-energy tensor. One then constructs a stress-energy tensor on the classical background by some means, denoted here as $\epsilon\,{_{(0)}\quantt}$, and attempts to solve ${_{(\hbar)}G_{\mu\nu}}= 8\pi\, {_{(0)}\quantt}$ for $h_{\mu\nu}$. This is the \emph{back-reaction problem}.

\subsection{Developments in semi-classical black hole research}

Perhaps one of the most amazing predictions to come out of
semi-classical theory is the evaporation of black holes. The
calculation was first carried out in Hawking's paper
\emph{``Particle creation by black holes''} in 1975
\cite{ref:hawkevap}. Although this is now an old result, given its importance it is
appropriate to review it in this section.

\begin{figure}[h!t]
\begin{center}
\includegraphics[bb=0 0 298 357, clip, scale=0.4, keepaspectratio=true]{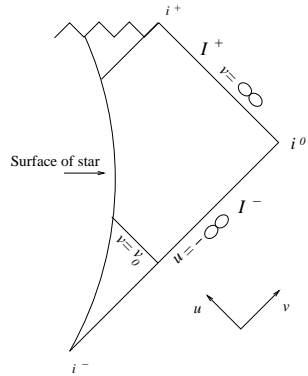}
\caption{{\small Penrose diagram depicting the surface of a
collapsing star.}} \label{fig:pencoll}
\end{center}
\end{figure}

We will study this effect in the Schwarzschild black hole
(\ref{eq:schwline}) in the forms given by (\ref{eq:inEF}) and (\ref{eq:outEF}). We
will also be considering Hawking's original argument where he
considered a collapsing star with asymptotically flat regions.
The Penrose diagram for the collapsing star is shown in figure
\ref{fig:pencoll}. Any direct interaction of the quantum particles with the material making up the stellar body will be ignored since the gravitational field of the star is the main ingredient for this effect and not the star itself.

For simplicity a null scalar field will be used to illustrate
this effect. At $\mathcal{I}^{-}$, there are no particles present
and the state will be denoted as $\left|0_{in}\right>$. It will
be found that, at $\mathcal{I}^{+}$, this state corresponds to a
state with a \emph{non-zero} particle content. This is due to the
presence of the non-trivial gravitational field.

Since $\mathcal{I}^{-}$ corresponds to approximately Minkowski
space-time, we can define modes there with frequency $\omega^{\prime}$ and
therefore the solutions to (\ref{eq:flatwaveeq}) in Schwarzschild space-time may be written as
\begin{equation}
\varphi^{\prime}_{\omega^{\prime},v}=f^{\prime}_{\omega^{\prime}}(\mathbf{x}) e^{-i\omega^{\prime} v} \:, \label{eq:farmodes}
\end{equation}
i.e. the standard particle states of flat space-time. The field
$\phi$ is written as before, with appropriate
creation and annihilation operators satisfying
$a_{\omega}\left|0_{in}\right>=0$: 
\begin{equation}
\phi=\int\left[a^{\prime}_{\omega^{\prime}}f^{\prime}_{\omega^{\prime}}(\mathbf{x})e^{-i\omega^{\prime} v} + a^{\prime\dagger}_{\omega^{\prime}}f^{\prime *}_{\omega^{\prime}}(\mathbf{x})e^{i\omega^{\prime} v}\right] d\omega^{\prime}\:. \label{eq:scalarin}
\end{equation}

(Note that in this argument we only
have inward propagating waves so only inward components are
necessary here.)

At $\mathcal{I}^{+}$ the space-time is also Minkowski. We can
define modes there as well, which in terms of the null coordinate $u$ have the form:
\begin{equation}
\varphi_{\omega,u} = f(\mathbf{x})_{\omega} e^{-i\omega u} \:, 
\label{eq:futurefarmodes}
\end{equation}
which again correspond to the usual flat space-time modes with
positive frequency. These are outgoing modes.

Now, we can express $\varphi$ in terms of $\varphi^{\prime}$ utilizing a similar transformation as in (\ref{eq:wavetransf}):
\begin{equation}
\varphi=\int \left[\alpha^{*}_{\omega^{\prime}\omega}\varphi^{\prime}-\beta_{\omega^{\prime}\omega} \varphi^{\prime *}\right]dv \,. \label{eq:out_as_fun_in}
\end{equation}
As in the flat space-time case, should the second Bogoliubov coefficient not vanish, there will be a disagreement in particle content between an ``in'' observer and an ``out'' observer.

Let us consider the transformation between the in and out observers. For a single outgoing mode we have:
\begin{equation}
 e^{-i\omega u}=e^{-i \omega\left(v-2r^{*}\right)} =e^{-i \omega \left[v-2r -4M \ln\left(\frac{r}{2M}-1\right)\right]}.
\end{equation}
Consider now tracing back the path of a particle (outgoing) as it skims the horizon, near $r=2M$. If the star is collapsing, the level surface representing the stellar boundary (assumed to be near $r=2M$) changes with time as the particle traverses through the star. The equation describing this surface is given by $r\approx 2M+\kappa_{0}(v-v_{0})+ \mathcal{O}(v-v_{0})^{2}$. Adopting the eikonal approximation, and assuming most of the change in the eikonal takes place near this surface we get
\begin{equation}
 \omega u \approx \omega \left[v -4M +2 \kappa_{0}(v-v_{0}) + ... -4M \ln \left(\frac{\kappa_{0}(v_{0}-v)}{2M}+ ...\right)\right]\: , \label{eq:outeik} 
\end{equation}
with $v < v_{0}$.

\begin{figure}[h!t]
\begin{center}
\includegraphics[bb=0 0 329 420, clip, scale=0.4, keepaspectratio=true]{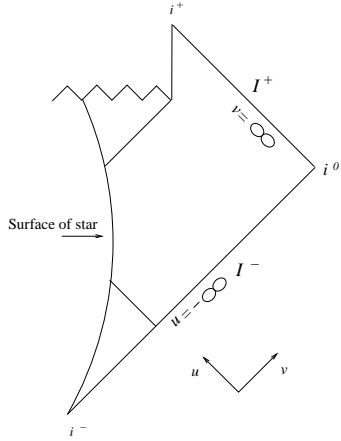}
\caption{{\small Penrose diagram of stellar collapse taking into
account subsequent black hole evaporation.}} \label{fig:penevap}
\end{center}
\end{figure}

Concentrating on the dominant part of the eikonal, we can see that the ingoing eikonal (which we denote as $\Omega$) corresponding to the outward eikonal (\ref{eq:outeik}) is given by:
\begin{equation}
 \Omega(v) \approx -4M\omega \ln \left((v_{0}-v)\right) + \mbox{constant}\:,
\end{equation}
the constant shift being irrelevant. Comparing this expression with (\ref{eq:out_as_fun_in}) it may be seen that, as a Fourier transform, $\varphi_{\omega,u}$ has both $e^{-i\omega^{\prime}v}$ and $e^{+i\omega^{\prime}v}$ components and therefore the corresponding $\beta_{\omega^{\prime}\omega}$ \emph{does not vanish} in the transformation. There are therefore out particles even in the absence of in particles.
The conformal diagram corresponding to the collapse into, and subsequent evaporation of, a black hole is shown in figure \ref{fig:penevap}. A much more detailed calculation, which includes a derivation of the actual particle spectrum, may be found in \cite{ref:traschen}.

This evaporation issue is related to what is sometimes known as the information loss problem where ``information'' falling into a purely classical black hole is presumably lost due to the fact that a black hole is described solely by its mass, charge and angular momentum and therefore the evolution is not unitary. It is thought that if the resulting black hole radiation spectrum is not truly thermal then the information may re-emerge from the black hole in this way. Later it will be shown that a full theory of quantum gravity may resolve this in another way.

Since the Hawking result, much work has been done in semi-classical gravity, major advancements having been made in the 1980s. Much of the early works concentrated on methods to obtain sensible, convergent quantities within semi-classical theories. The interested reader is referred to \cite{ref:BandD} and references therein. In the following we limit our survey to some issues which directly involve the back-reaction problem as the literature and topics in the field of semi-classical gravity are almost as vast as in the classical theory making anything resembling a complete coverage nearly impossible.

More recently, semi-classical black hole research has focused on modeling the perturbations on the classical background geometry due to quantum fields and their fluctuations. Study of the latter effect makes up the arena of stochastic gravity. These problems require computation of the renormalized stress-energy tensor on the background, which is then to be used as a source for the metric perturbations. 

In the black hole context, Hiscock, Larson and Anderson have calculated the back-reaction effects of scalar, spinor and vector fields inside a Schwarzschild black hole's event horizon \cite{ref:hla}. To construct $\quantt$\footnote{From now on in this section we drop the $(0)$ subscript in front of $\quantt$ and it is understood that $\quantt$ refers to the first-order quantum correction to the stress-energy tensor and is constructed with the zeroth-order (classical) background metric.} they utilized various approximation schemes developed in the literature \cite{ref:pageapprox} - \cite{ref:andapprox} and the DeWitt-Schwinger expansion \cite{ref:dewschwing}. They studied the quantum field's back reaction on the anisotropy of the interior as well as the first-order correction to the Kretschmann scalar. In summary they found the following: Spinors and minimal and conformally coupled scalars tended to decrease the anisotropy as the singularity was approached whereas vector fields tended to increase the anisotropy. Regarding the effect on the Kretschmann scalar, it was found that the minimally coupled massive scalar field and spinor fields tend to slow down the rate of increase of curvature as one approaches the singularity whereas other couplings and fields tended to increase the rate of curvature growth. 

A similar problem was studied for the case of cylindrical black holes (or black strings) in \cite{ref:blackstringback} utilizing the stress-tensor approximation in \cite{ref:pageapprox} for quantum scalar fields. It was found there that the conformally coupled scalar also tended to increase the growth of curvature near the horizon. In this case, utilizing the cylindrical version of the metric (\ref{eq:onenulltopo}) as a background space-time, with cylindrical (rather than one null coordinate) it was found that
\begin{equation}
 \delta K \approx \left[\frac{3}{10} \frac{\alpha^{4}}{\pi} -2 \frac{\alpha^5}{\pi} \left(\frac{2}{M}\right)^{1/3} \left(T - \frac{(4M)^{1/3}}{\alpha}\right)\right], \label{eq:kretschpert}
\end{equation}
where $\delta K$ is the perturbation on the background Kretschmann scalar. Here $T$ is the interior time coordinate (corresponding to the radial coordinate outside the black string) and the horizon is located at $T=\frac{(4M)^{1/3}}{\alpha}$.

Of course, the above results are only valid insofar as the perturbation is valid and the results cannot be extrapolated right down to the singularity.

In spherical black holes an enormous amount of work has been done in calculating field expectation values and stress-energy tensors of various fields and couplings. There has also been much work in trying to produce the $A/4$ entropy of black holes from semi-classical theory. The amount of work in this fascinating area is too large to even begin reviewing here. We refer the interested reader to \cite{ref:entreview2} and references therein along with the books \cite{ref:BandD} \cite{ref:waldbhthermo}.

For cylindrical black holes the amount of work is much more modest. Very briefly, in the context of the cylindrical black holes, Piedra and de Oca have studied the quantization of massive scalar and spinor fields over static black string backgrounds \cite{ref:pdo1} \cite{ref:pdo2}. They have calulated $\quantt$ up to second order in the inverse mass value. Dias and Lemos have studied magnetic strings in anti-de Sitter general relativity \cite{ref:adsmagstring} and the scalar expectation value, $\left<\phi^{2}\right>$, has been computed in \cite{ref:scalexp}. 

Interestingly, critical behavor has also been studied in the context of semi-classical gravity. For technical reasons, much of this work has been done in two dimensional tensor-dilaton gravity as $\quantt$ may be determined from the trace anomaly along with the mild assumption of conservation. Ayal and Piran, for example, obtained a critical scaling exponent of $\eta \approx 0.409$ \cite{ref:aypir}. A slightly different model, utilizing a conformally coupled scalar, was analyzed in \cite{ref:stromthor} and a critical exponent of $0.5$ was found. In \cite{ref:pbp} yet another model was employed which allows one to turn the quantum effects on and off. A critical scaling exponent of $\eta \approx 0.53$ was found in this study. In \cite{ref:bradott} the authors calculated the quantum stress tensor on a classical background spacetime with perfect fluid source. The quantum effects then were treated as perturbations of the classical fluid gravitating system. They found that a mass gap exists when $\eta \geq 0.5$ so that there is a minimum size to the black holes formed. The case $\eta < 0.5$ could not be studied as the semi-classical approximation breaks down in that regime.

In stochastic gravity one takes the level of quantum approximation one step further, considering the effects of the field fluctuations. Given limited space we cannot cover stochastic gravity here. The interested reader is referred to the reviews \cite{ref:stoch1} - \cite{ref:stochfinal} and references therein.

\section{Quantum Black Hole Research}
\subsection{Quantum gravity}
Attempts to quantize gravity go almost as far back as the dawn of
quantum mechanics. One of the earliest arguments for the
quantization of gravity, in fact almost as old as general
relativity itself, is that if $T_{\mu\nu}$ is inherently quantum,
then so it should be with the gravitational field which it produces
\cite{ref:hintqg}. However it can be argued that this is not
necessarily the case \cite{ref:rosenfeld}. Although it is now generally believed that the gravitational field must be quantized in some way, there is still
some debate on this necessity.

In the early days of quantum theory, the first person to realize that there would be serious problems applying those techniques to gravity seems to have been Matevi Bronstein. Bronstein had the insight to deduce that quantum theory could not be applied in any obvious way to a theory that was background independent. It was possible, according to Bronstein, that the ordinary notions of space and time would have to be abandoned \cite{ref:bron}. Other pioneers in the field included P. Bergmann and P. Dirac.

After quantum field theoretic techniques were sufficiently developed, an
obvious approach to quantizing gravity was implemented as a simple background
expansion:
\begin{equation}
 g_{\mu\nu}= \eta_{\mu\nu} + \epsilon h_{\mu\nu} + \mathcal{O}(\epsilon^{2})\:. \label{metexp}
\end{equation}
As is now well known, treating the perturbations as fields on the
background metric ($\eta_{\mu\nu}$) yields a non-renormalizable
quantum field theory with divergences commencing at the one-loop
level for gravity with matter couplings. 

In the early 60s, Feynman, working at tree-level, computed transition amplitudes and demonstrated that reasonable results are obtained. This gave hope for this line of quantum gravity research. However, he noted that at loop level problems began to arise which required the introduction of Faddeev-Popov ghosts by DeWitt \cite{ref:oneloopprobs1}-\cite{ref:oneloopprobs4}.

In the 70s it became generally accepted that gravity coupled to matter will be non-renormalizable. It was, however, found that one could add a spin $\frac{3}{2}$ particle to general relativity which yields a theory finite at two-loops. Thus began the field of study known as supergravity \cite{ref:superg}.

Finally, it was shown explicitly in 1986, by Goroff and
Sagnoti that, at two-loop order, finite S matrix elements could
be attained if the gravity action contained a counter-term of the
form\footnote{At one loop order the divergence possesses terms
proportional to $R^{2}$ and $R_{\mu\nu}R^{\mu\nu}$ and is
therefore finite in the absence of matter.} \cite{ref:GS}:
\begin{equation}
\mathcal{L}_{ct}=\frac{1}{(d-4)}\frac{209}{2880
(16\pi)^{2}}\sqrt{-g}R^{\alpha\beta}_{\;\;\;\,\gamma\delta}\,R^{\gamma\delta}_{\;\;\;\,\mu\nu}\,
R^{\mu\nu}_{\;\;\;\,\alpha\beta}\:, \label{eq:GS}
\end{equation}
with $d$ the effective regularized dimension of the space-time, thus explicitly showing the divergent properties at two-loop level.

On the canonical side, DeWitt in 1967 publishes what was originally thought of as the ``Einstein-Schr\"{o}dinger equation'' also known as the Wheeler-DeWitt equation \cite{ref:wdw}. Some argued at the time that the problem of quantizing the gravitational field had been solved. 

\begin{figure}[h!t]
\begin{center}
\includegraphics[bb=0 0 903 354, clip, scale=0.35, keepaspectratio=true]{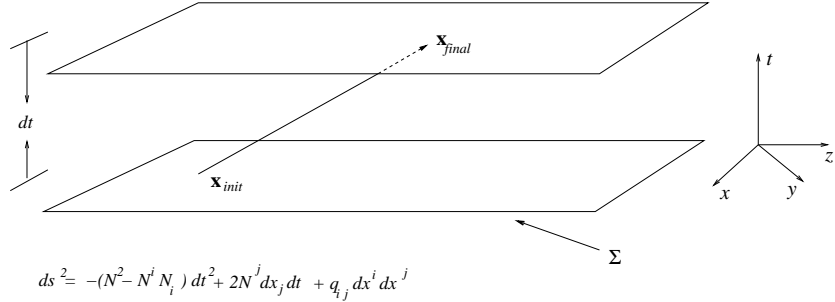}
\caption{{\small ADM decomposition of space-time into ``space'' and
``time''.}} \label{fig:adm}
\end{center}
\end{figure}

This Hamiltonian approach begins with the familiar ADM decomposition of space-time, as illustrated in figure \ref{fig:adm}. In the figure $q_{ij}$ is the metric of the three-surface $\Sigma$ and $N$ and $N^{j}$ are the usual lapse function, and shift vector associated with the ADM decomposition.

With this decomposition, the gravitational action may be written as:
\begin{eqnarray}
 I&=&\frac{1}{16\pi} \int dt \int_{\Sigma}dx^{3} \left[\Pi^{ab}\dot{q}_{ab}\right.\nonumber \\
 &&\left.+2N_{b} \nabla^{(3)}_{a} \left(q^{-1/2} \Pi^{ab}\right) +Nq^{1/2}\left(R^{(3)} -q^{-1}\Pi_{cd}\Pi^{cd} +\frac{1}{2} q^{-1}\Pi^{2}\right)\right], \label{eq:ADMact}
\end{eqnarray}
with $\Pi:=\Pi^{ab}q_{ab}$, $\Pi^{ab}:=q^{-1/2}\left[K^{ab}-K\,q^{ab}\right]$, $K:=K^{ab}q_{ab}$. $K_{ab}$ is the extrinsic curvature tensor and over-dots denote differentiation with respect to the slicing time, $t$. In this scheme, the field variable is $q^{ab}$ and its conjugate momentum is $\Pi^{ab}$. The equations of motion go over to
 \begin{eqnarray}
  2\nabla^{(3)}_{i}\left[q^{-1/2}\Pi^{ij}\right]=0 =:-V^{j}\:, \nonumber \\
  q^{1/2}\left[R^{(3)}-q^{-1} \Pi_{ab}\Pi^{ab} + \frac{1}{2}q^{-1}\Pi^{2}\right]=0=:-S\:. \nonumber
 \end{eqnarray}
The action leads to a Hamiltonian density:
 \begin{eqnarray}
  \mathcal{H}_{G}=\frac{1}{16\pi}\Pi^{ab}\dot{q}_{ab} -\mathcal{L}_{G} =N_{b}V^{b}+N\:S\;, \nonumber
 \end{eqnarray}
 and the following symplectic structure:
 \begin{eqnarray}
\left\{\Pi^{ab}(\mathbf{x}),\, q_{cd}(\mathbf{y})\right\}=16\pi \delta^{a}_{\,(c}\delta^{b}_{\;d)} \delta(\mathbf{x},\,\mathbf{y}), \nonumber \\
  \left\{\Pi^{ab}(\mathbf{x}),\,\Pi^{cd}(\mathbf{y})\right\} = 0 = \left\{q_{ab}(\mathbf{x}),\, q_{cd}(\mathbf{y})\right\}\:, \nonumber 
 \end{eqnarray}
 which after quantization leads to the famous Wheeler-DeWitt equation:
 \begin{eqnarray}
\left[\left[-q_{ab}q_{cd}+\frac{1}{2}q_{ac}q_{bd}\right]\frac{\delta}{\delta q_{ac}} \frac{\delta}{\delta q_{bd}} +q^{1/2}R^{(3)}\right] \Psi(q)=0. \label{eq:WdW} 
 \end{eqnarray}
The Wheeler-DeWitt formulation suffers from some problems. The configuration field (3-metric) does not appear as a gauge field. As well, there are inconsistencies with certain transition probabilities in the path-integral version.

There are several other candidate theories of quantum gravity. These include the sum-over-Euclidean geometries developed by Hawking \cite{ref:hawkbook} and its Lorentzian counterpart, the causal set approach of Sorkin \cite{ref:causalset1} \cite{ref:causalset2}, dynamical triangulations \cite{ref:dytri}, and other theories, including loop quantum gravity. Interestingly, the causal set approach predicts the existence of a small positive cosmological constant of the order of that required to provide the observed acceleration of the universe.

\subsubsection{The loop quantum gravity program in brief}
Loop quantum gravity provides a promising quantization scheme for general relativity. There is a Hamiltonian approach and a covariant approach, yielding a spin-foam model, so named due to the resemblance of the Feynman diagram analogs to a foam of bubbles. We will concentrate here on the Hamiltonian approach, which is perhaps a bit more perspicuous. For a nice review of the spin-foam approach, the reader is referred to \cite{ref:spinfoam}. 

It was noted by Ashtekar \cite{ref:ashvar} that general relativity can be very neatly reformulated in terms of a densitized triad, $E_{j}^{b}$ instead of the metric:
\begin{eqnarray}
  q\,q^{ab}=E_{\;i}^{a}E_{j}^{b}\delta^{ij} \nonumber
 \end{eqnarray}
 with 
 \begin{eqnarray}
  E^{a}_{\;i}:=\frac{1}{2} \epsilon^{abc} \epsilon_{ijk}e^{j}_{\;b}e^{k}_{\;c},\;\;\; \mbox{where}\;\;\; q_{ab}=e^{i}_{\;a}e^{j}_{\;b} \delta_{ij}. \nonumber
 \end{eqnarray}
This is sometimes known as the phase-space representation and the indices $i,\,j$ etc. denote the orthonormal components. In terms of the new variables, the ADM action (\ref{eq:ADMact}) may be written as
\begin{eqnarray}
I=\frac{1}{8\pi} \int dt\int_{\Sigma} \left[E^{a}_{\;i} \dot{K}^{i}_{\;a} -N_{b}V^{b} -NS -N^{i} \epsilon_{ijk}E^{aj}K^{k}_{\;a} \right]d^{3}x\;, \label{eq:ashact}
 \end{eqnarray}
with $K^{i}_{\;a}:=\frac{1}{\sqrt{\det (E)}} K_{ab} E^{b}_{\;j} \delta^{ij}$ . In this scheme, the canonically conjugate variables are $E^{a}_{\;i}$ and ${K}^{i}_{\;a}$.

The symmetry group in $\Sigma$ is $SO(3)$. We can write $K^{i}_{\;a}$ in terms of the fiducial $so(3)$ connection\footnote{The fiducial connection is that yielded by the solution of Cartan's structural equation, $\partial_{[a}e^{i}_{\;b]}+ \epsilon^{i}_{\;jk}\Gamma^{j}_{\;[a}e^{k}_{\;b]}=0$, where $e^{i}_{\;a}$ is the standard (non-densitized) triad, $q_{ab}=e^{i}_{\;a}e^{j}_{\;b}\delta_{ij}$.} on $\Sigma$, $\Gamma^{i}_{\;a}$:
 \begin{eqnarray}
  \gamma K^{i}_{\;a}=A^{i}_{\;a} - \Gamma^{i}_{\;a}\;, \nonumber
 \end{eqnarray}
where $\gamma$ is known as the \emph{Immirzi parameter}. The ``modified'' connection $A^{i}_{\;a}$ can be defined by this equation. In terms of this new connection the action may be written as
\begin{eqnarray}
 I&=&\frac{1}{8\pi} \int dt\int_{\Sigma} \left[E^{a}_{\;i} \dot{A}^{i}_{\;a} -N_{b}V_{b} -NS -N^{i} G_{i} \right]d^{3}x\;, \label{eq:Aact}\\
  G_{i}&:=&\partial_{a}E^{a}_{\;i} +\epsilon_{ij}^{\;\;\;k}A^{j}_{\;a}E^{a}_{\;k}\:, \nonumber 
\end{eqnarray}
with symplectic structure:
\begin{eqnarray}
 \left\{E^{a}_{\;j}(\mathbf{x}),\,A^{i}_{\;b}(\mathbf{y})\right\}=8\pi \gamma \delta^{a}_{\;b} \delta^{i}_{\;j}\,\delta(\mathbf{x},\mathbf{y})\:, \nonumber \\
 \left\{E^{a}_{\;i}(\mathbf{x}),\, E^{b}_{\;j}(\mathbf{y})\right\}=\left\{A^{j}_{\;a}(\mathbf{x}),\,A^{i}_{\;b}(\mathbf{y})\right\}=0\:. \nonumber
 \end{eqnarray}
The above formulae do not distinguish between $SO(3)$ and $SU(2)$ and both these groups possess the same algebra so it is customary to work in $SU(2)$, the indices $i,\,j$ now coupling quantities to the $su(2)$ algebra. The quantization of this system yields the canonical version of loop quantum gravity. $E^{a}_{\;i}$ and $A^{i}_{\;a}$ are the \emph{Ashtekar - Barbero variables}. The quantum versions of the equations of motion yield the \emph{quantum Einstein equations}:
\begin{subequations}
\begin{align}
  \hat{G}_{i}\psistate =&\widehat{D_{a}E^{a}_{\;i}}\left|\Psi\right> =0\:, \label{eq:qe1} \\
  \hat{V}_{a}\psistate =& \left[\widehat{E^{a}_{\;i}F^{i}_{\;ab}} -(1-\gamma^{2}) \widehat{K^{i}_{\;b}G_{i}}\right]\psistate =0\:, \label{eq:qe2}  \\
  \hat{S}\psistate =&\left[\frac{1}{\sqrt{\det(E)}} \widehat{E^{a}_{\;i}E^{b}_{\;j} \epsilon^{ij}_{\;\;\;k}F^{k}_{\;ab}} -2(1+\gamma^{2}) \widehat{K^{i}_{\;[a}K^{j}_{\;b]}} \right]\psistate=0\:, \label{eq:qe3}  
 \end{align}
 \end{subequations}
 with
 \begin{align}
 F^{i}_{\;ab}:=& \partial_{a}A^{i}_{\;b} -\partial_{b} A^{i}_{\;a} +\epsilon^{i}_{\;\;jk}A^{j}_{\;a}A^{k}_{\;b}\:, \nonumber \\
D_{a}v_{i}:=&\partial_{a}v_{i} -\epsilon_{ijk}A^{j}_{a}v^{k}\;. \nonumber
 \end{align}
Note that now there is a constraint equation for $G_{i}$ (the Gauss constraint). This reflects the fact the triad possesses a rotational freedom; one can choose different frames locally by rotating the triad. This redundancy is eliminated in the new Gauss constraint. To the above system one could add matter couplings by supplementing the action with a matter term and quantizing appropriately. The problems associated with (\ref{eq:WdW}) are not present in this representation.

Before continuing, we shall make a few comments about this quantization scheme:
 \begin{enumerate}
  \item The scheme is background independent and respects diffeomorphism invariance. The choice of time slicing is arbitrary and does not affect the physics.   
  \item A superpartner can be accommodated and therefore supersymmetry can be incorporated. This has been done \cite{ref:superLQG}.
  \item Instead of the Einstein-Hilbert action, one can accommodate geometric actions made up of arbitrary curvature invariants. The scheme is generally similar to that outlined above.
  \item Higher dimensions can be accommodated.
 \end{enumerate}
It should be noted that 2, 3, and 4 are \emph{not required} but simply can be accommodated.

\begin{figure}[ht]
\begin{center}
\includegraphics[bb=0 0 574 390, clip, scale=0.40, keepaspectratio=true]{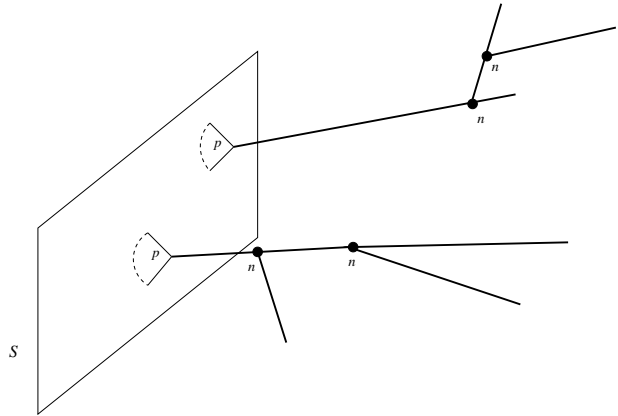}
\caption{{\small A gravitational spin-network endowing a surface $S$
with area and geometry. The spin-network punctures the surface
$\mathcal{S}$ at $p$. The surface is flat everywhere except at the
punctures. There, the spin-network can be pictured as ``tugging'' on
the surface, endowing it with geometry and introducing a local
angular defect on the surface. The $n$'s are the nodes, which are
associated with volumes.}} \label{fig:spinpierce}
\end{center}
\end{figure}

What is of interest is the holonomy of $A$ as it is transported around what are known as \emph{spin-networks}, (first introduced by Penrose \cite{ref:penspin} and utilized early in loop quantum gravity by Jacobson and Smolin \cite{ref:smolspin}) and the state vectors $\psistate$ which are functions of this holonomy. The concept of time evolution is now encoded in terms of how the interrelationship of the network, which describes space, evolves. The details are beyond the scope of this manuscript but the interested reader may find them in \cite{ref:lqgrev1}, \cite{ref:lqgrev2}, \cite{ref:lqgrevn}. What is of particular interest in the context of black hole research is that this theory predicts that on the small scale, space is \emph{discrete}!\footnote{It should be emphasized that this is a \emph{prediction} of the theory and is not put in ``by hand''.} Classically, the area may be constructed out of the triad via
\begin{equation}
 A(\mathcal{S})=\int_{\mathcal{S}} \sqrt{n^{a}E^{i}_{a}n_{b}E^{b}_{i}}\,d^{2}s\:,
\end{equation}
with the $n$ vectors denoting normals to the 2-surface  $\mathcal{S}$. To go over to the quantum theory one replaces the classical triad with the corresponding quantum operator. The picture that arises is that each fiber of the spin-network that pierces a surface $\mathcal{S}$ endows it with a certain amount of area and geometry, via the introduction of an angular defect (see figure \ref{fig:spinpierce}).

The triad possesses the following spectrum when acting on the state functions of loop quantum gravity:
 \begin{eqnarray}
 \hat{E}_{i}(S_{I})\hat{E}^{i}(S_{I})\,\Phi^{j}\left(h_{e}[A]\right) = (8\pi l_{p}^{2}\gamma)^{2} \left[j_{p}(j_{p}+1)\right] \Phi^{j}\left(h_{e}[A]\right) \;,\nonumber
\end{eqnarray}
where $\Phi^{j}\left(h_{e}[A]\right)$ are state functions which depend on the holonomy of $A$, $h_{e}[A]$, along an edge $e$ of the spin-network. The $j_{p}$ are half-integers and $l_{p}$ is the Planck length. Therefore, for the area operator\footnote{We are making an assumption here regarding how the spin-network pierces the surface $\mathcal{S}$. The general case yields eigenvalues which are slightly more complicated than (\ref{eq:areaspec}).} :
  \begin{eqnarray}
   \widehat{Area}_{S}\psistate=8\pi l_{p}^{2} \gamma \sum_{p}\sqrt{j_{p}(j_{p}+1)}\psistate\:. \label{eq:areaspec}
  \end{eqnarray}
Notice that the eigenvalues of area are \emph{discrete}!

One can also construct the classical volume utilizing the triad:
\begin{equation}
 V=\int_{M} \sqrt{\left| \frac{1}{3!} \epsilon_{abc}E^{a}_{i}E^{b}_{j}E^{c}_{k} \epsilon^{ijk}\right|}\,d^{3}x\:.
\end{equation}
This can be replaced by its quantum analog and volume eigenvalues can be calculated. The results are somewhat complicated and we omit them here. However, the volume is also discrete.

We will see below that these two operators are of extreme importance in loop quantum gravity black hole research.

\subsection{Developments in loop quantum black hole research}
There are a number of results regarding black holes in loop quantum gravity. We shall concentrate here on what are arguably the two most significant results; namely the source of black hole entropy and the resolution of the singularity problem. These are of importance because it has long been believed that any viable theory of quantum gravity should explain where the enormous entropy of a black hole comes from and it should also eliminate singularities present in the classical theory.

\subsubsection{Black hole entropy}
The subject of black hole entropy has been one of intense interest ever since Bekenstein's calculations \cite{ref:bek}. Many methods have since been utilized to calculate the entropy (see
\cite{ref:entreview}, \cite{ref:entreview2} and references therein
for excellent reviews of the subject). One
belief is that the source of this entropy is strictly gravitational in origin.
That is, one should be able to define microstates in a full
quantum theory of gravity which, when counted, yields the
correct entropy law. This has been done within the framework of loop quantum gravity.

\begin{figure}[h!t]
\begin{center}
\includegraphics[bb=0 0 513 395, clip, scale=0.45, keepaspectratio=true]{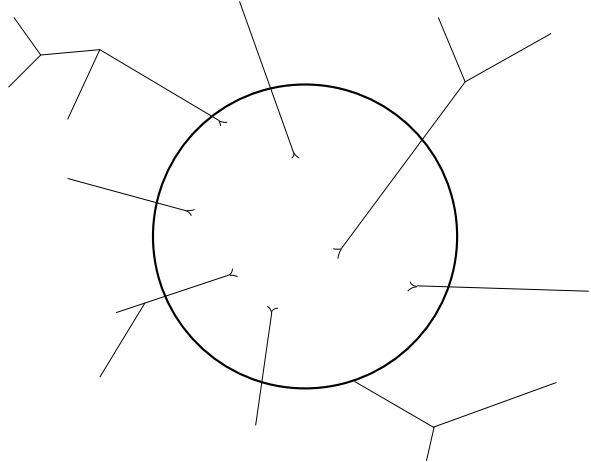}
\caption{{\small A gravitational spin-network giving a spherical
black hole horizon its geometry and area.}}
\label{fig:puncturedsurface}
\end{center}
\end{figure}

The basic idea is as follows: The gravitational spin-network pierces the surface corresponding to the horizon of the black hole. As described above, this endows the surface with area and geometry (see figure \ref{fig:puncturedsurface}).
The entropy is given by the logarithm of the number of loop quantum gravity states that give the surface a fixed area, $a_{0}$. This counting is non-trivial as for a black hole of reasonable size there could be an enormous number of punctures, with various values of $j_{p}$. The total area is given by summing up all the contributions from all of the punctures. The total area is therefore given by\footnote{We are making a similar assumption here as in the previous footnote.}
\begin{equation}
 a_{0}=8\pi l_{p}^{2}\gamma\sum_{p}\sqrt{j_{p}(j_{p}+1)}\:, \label{eq:horizarea}
\end{equation}
which is obviously the sum of eigenvalues associated with the edges puncturing the surface. Notice that this formulation yields the entropy as a function of $\gamma$. The quantity $\gamma$ therefore has to be set by other means or else it can be set by demanding that the leading order term in the entropy calculation agrees with the Hawking-Bekenstein result of $a_{0}/4$. Several studies have set the Immirzi parameter in this way \cite{ref:entrev2} - \cite{ref:jacobson}.

There is another subtlety which complicates the expression which is topological in origin. To illustrate this we begin by noting that the action giving rise to the surface states (and thus the surface Hilbert space) is a Chern-Simons action:
\begin{equation}
I_{|\partial M}=-\frac{i\,a_{0}}{32\pi \gamma (g-1)}\int_{\partial M} \mbox{Tr}\left[A \wedge dA +\frac{2}{3} A \wedge A \wedge A\right]\:, \label{eq:csact}
\end{equation}
where the trace of the $SU(2)$ connection, $A^{i}_{\;a}$ is taken over the $su(2)$ indices. The form of this surface term was first calculated in the pioneering work of \cite{ref:ACK}. Later this was generalized to arbitrary genus $g$ surfaces, as above \cite{ref:KBD}. This term arises from the fact that at the inner boundary (the isolated horizon \cite{ref:isohoriz}), the triad and the connection cannot be fixed independently and are actually related (hence only the connection appears in (\ref{eq:csact})). The isolated horizon boundary conditions reduce the degrees of freedom and the above action can be written in terms of a $U(1)$ connection \cite{ref:ACK}:
\begin{equation}
 I_{|\partial M}=\frac{i\,a_{0}}{16\pi^{2} (1-g)}\int_{\partial M} W \wedge dW \:, \label{eq:u1form}
\end{equation}
where $W$ represents $U(1)$ connections on the boundary surface, which are restricted by
the value of the bulk $SU(2)$ connection penetrating the surface at that particular point on the horizon. Finally one can construct the symplectic structure on the boundary as was done in \cite{ref:ACK}:
\begin{equation}
 \Omega_{|\partial {M}\,grav}(.., \, ..)= k \oint_{\partial M} \delta W \wedge \delta^{\prime}W\:, \label{eq:gravsymp}
\end{equation}
with $k:=\frac{a_{0}}{4\pi(g-1)\gamma}$ (an integer) known as the Chern-level and with $\delta W$ and $\delta^{\prime}W$ tangent vectors in the space of $U(1)$ connections defined on the horizon.
Note that we now have a topological $U(1)$ theory on the boundary. The number degrees of freedom are related to the number of topologically independent closed paths one can construct on the punctured surface.

In the case of a surface with spherical topology $S^{2}$, one can place punctures on the sphere and then define a closed path around each puncture. This would seemingly yield a $2N$ dimensional phase-space as each closed path (cycle) also has a conjugate open path associated with it (chain) (roughly speaking each cycle represents a configuration variable that must have a conjugate momentum, represented by chains). However, these are not all independent degrees of freedom. On a sphere, note that going around the $N$-th puncture is the same as going around all the other punctures but in the opposite direction. If the cycles are denoted as $\eta_{i}$,  this may be expressed as:
\begin{equation}
 \eta_{1}\cdot\eta_{2}\cdot ... \cdot \eta_{N-1}=\eta_{N}^{-1}\:, \label{eq:etarel}
\end{equation}
Therefore, the topology reduces the number of independent degrees of freedom. 

When one quantizes the system, quantum states $\psi_{m}$ are obtained for integers
$m=(m_{1},..,m_{N-1})$ with $m_{i}\in\left\{1, .., k\right\}$
\cite{ref:ABK}. Note that in this sense, the integers $m_{i}$ play a role similar to
the magnetic quantum number in ordinary quantum mechanics. The condition
(\ref{eq:etarel}) gives rise to a constraint:
\begin{equation}
 m_{1}+...+m_{N-1}=-m_{N}\;. \label{eq:projconst}
\end{equation}
This restriction is the quantum analogue of the Gauss-Bonnet theorem for a sphere.
Note that one now has $N$ generators and one constraint. Thus, for a spherical horizon, states can be
labeled with $m=(m_{1},..,m_{N})$ subject to constraint (\ref{eq:projconst}). In other words, \emph{not all} states that yield classical area $a_{0}$ are allowable. Only the ones meeting the condition (\ref{eq:projconst}) are to be counted. The details of the counting may be found in \cite{ref:correview} and references therein. Only the results will be cited here.

A very careful numerical counting of acceptable states for a spherical horizon was performed by Corichi et al in \cite{ref:entrev6}. Those authors performed the counting with the projection constraint (\ref{eq:projconst}) and without considering it. They found the following: \emph{Without} considering the projection constraint they found that the entropy obeys the $a_{0}/4$ law, provided the Immirzi parameter is set equal to $\gamma\approx 0.274$. This is the same value found in \cite{ref:entrev3} and \cite{ref:correview}. \emph{With} the projection constraint it was found that the number of acceptable states is reduced in such a way that does not involve $\gamma$. The result is 
\begin{equation}
 S=\frac{a_{0}}{4}-\frac{1}{2}\ln(a_{0})\:. \label{eq:sphereent}
\end{equation}

For the case of a genus $g$ surface, the situation is slightly more subtle as the paths around the punctures can also be related to paths around the genus holes of the surface. This yields a quantum Gauss-Bonnet theorem of the following type \cite{ref:KBD}:
\begin{equation}
 \eta_{g+1}\cdot\eta_{g+2}\cdot ... \cdot \eta_{g+N}= \eta_{1}\gamma_{1}\eta_{1}^{-1}\gamma_{1}^{-1} \cdot ... \cdot \eta_{g}\gamma_{g}\eta_{g}^{-1}\gamma_{g}^{-1}\:, \label{eq:torusrelg}
\end{equation}
where $\eta_{1}$ through $\eta_{g}$ denote the paths associated with the genus holes and $\eta_{g+1}$ to $\eta_{g+N}$ denote with the paths associated with the spin-network punctures. Utilizing this topological condition, the entropy of a $g > 1$ horizon is given by \cite{ref:KBD}
\begin{equation}
 S= \frac{a_{0}}{4} +(g-1)\left[ \ln (a_{0}) -\ln (4\pi\gamma (g-1))\right]\:, \label{eq:theentropy}
\end{equation}
provided that the Immirzi parameter is set to the same value as in the spherical case. Therefore, the same value of the Immirzi parameter yields the first-order $a_{0}/4$ term for all cases whereas the sub-leading term depends on topolgy and is independent of $\gamma$. This behavior is consistent with other, non-quantum gravity approaches to calculating black hole entropy of $g > 0$ horizons \cite{ref:govind} \cite{ref:vanzo}\cite{ref:mans} \cite{ref:liko}. 

An ambiguity exists for $g=1$ due to the decoupling of the triad and the connection at the horizon in this case. However, one may analytically extend (\ref{eq:theentropy}) to $g=1$ yielding an $a_{0}/4$ entropy \emph{without} logarithmic correction. This result is consistent with studies of $g=1$ horizon entropy utilizing non loop quantum gravity techniques \cite{ref:vanzo}, \cite{ref:liko}. The $g>1$ result however, is qualitatively different from the $g=0$ case, and therefore cannot be extended to reliably encompass the $g=0$ horizons. This is due to the non-trivial interplay between the spin-network punctures and the genus holes of the surface (note that the coefficient of the logarithmic correction in (\ref{eq:theentropy}) differs from (\ref{eq:sphereent}) by a factor of 2 for $g=0$).

\subsubsection{Removal of the classical singularity}
Another problem that a quantum theory of gravitation is expected to resolve is that of the singularities that exist in the classical theory. This is also related to the problem of information loss associated with black holes. This singularity issue has been studied in the framework of loop quantum gravity in the case of mini-superspace models. That is, models where the full system is first reduced to a mechanical system which consists of only the relevant degrees of freedom. One then quantizes this reduced system. This is done due to the technical difficulty involved when trying to work with the full theory. There are currently several black hole studies available within the symmetry reduced models \cite{ref:lqgsing1} - \cite{ref:AB} and here we shall be outlining the approach in \cite{ref:AB}. A related method was studied in \cite{ref:vander} where it was shown that a Nariai universe replaces the classical singularity.

The basic ideas are as follows: Construct an evolution equation utilizing the Hamiltonian constraint and check if the evolution remains finite at the point corresponding to the classical singularity. Also, one may compute operators
and their expectation values which encode the information about curvature and which classically diverge at the classical singularity. If they remain finite in the quantization the singularity is avoided in the quantum theory. 

We will focus on the most studied black hole in quantum gravity, the Schwarzschild black hole ($\Lambda=0=Q$), whose line element for $r < 2M$ can be written as:
\begin{equation}
ds^{2}= -\frac{dT}{\frac{2M}{T}-1} + \left(\frac{2M}{T}-1\right)dR^{2} +T^{2}\,d\theta^{2} + T^{2}\sin^{2}\theta\, d\varphi^{2}\:, \label{eq:tdomline}
\end{equation} 
where $T$ is the interior time coordinate (corresponding to $r$ in (\ref{eq:krnsol})) and $R$ is an interior spatial coordinate (corresponding to $t$ in (\ref{eq:krnsol})).

Recall that the conjugate variables in loop quantum gravity are the densitized triad and the modified $SU(2)$ connection. A pair is is constructed which respects the symmetry of the space, $\mathbb{R}\times SO(3)$  \cite{ref:AB}, \cite{ref:wittensansatz}:
\begin{subequations}
\begin{align}
A^{i}_{\;a}\tau_{i}\,dx^{a}=&c \tau_{3}\,dR +(a\tau_{1} +b \tau_{2})d\theta+(a\tau_{2}-b\tau_{1})\sin\theta\,d\varphi +\tau_{3}\cos\theta\,d\varphi\:, \label{eq:asym} \\
E^{a}_{\;i}\tau^{i} \partial_{a}=&p_{c}\tau_{3}\sin\theta \partial_{R} +(p_{a}\tau_{1} +p_{b}\tau_{2}) \sin\theta\,\partial_{\theta} +(p_{a}\tau_{2}-p_{b}\tau_{1}) \partial_{\varphi}\:, \label{eq:esym}
\end{align} 
\end{subequations}
where the $\tau_{i}$ denote the standard $su(2)$ basis. The quantities $a,\,b,\,c$ and $p_{a},\,p_{b},\,p_{c}$ are to be determined and act as conjugate ``position-momentum'' pairs. The classical analog of the Gauss constraint can be satisfied but not in a unique way. Any pair that satisfy:
\begin{equation}
ap_{b}-bp_{a}=0\:, \label{eq:cgconst}
\end{equation}
will satisfy the Gauss constraint \cite{ref:AB}. Therefore, it is useful to set $a=p_{a}=0$ in the sequel. There is still some residual gauge freedom but we shall not discuss it here.

The co-triad can also be constructed as:
\begin{equation}
 \omega_{a}^{\;i}\tau_{i}dx^{a}=\frac{\mbox{sgn}\,p_{c}\:|p_{b}|}{\sqrt{|p_{c}|}} \tau_{3} \,dR + \sqrt{|p_{c}|}\frac{p_{b}}{|p_{b}|} \tau_{2}\, d\theta - \sqrt{|p_{c}|}\frac{p_{b}}{|p_{b}|}\tau_{1} \sin\theta\,d\varphi\:. \label{eq:cotriad}
\end{equation}
By comparison of (\ref{eq:esym}) (or (\ref{eq:cotriad})) with (\ref{eq:tdomline}) one can see that the following identification may be made:
\begin{equation}
 p_{b} = \sqrt{T\left(2M-T\right)}\,,\;\;\;\; p_{c}=\pm T^{2}\:. \label{eq:pcomps}
\end{equation}
Therefore, the degeneracy in (\ref{eq:cotriad}) at $p_{b}=0 \neq p_{c}$ corresponds to the classical horizon whereas $p_{c}=0$ corresponds to the classical singularity.

Next a basis is defined in the Hilbert space, these are denoted as $\frac{1}{\sqrt{2}}\left[\left|\mu,\,\tau\right> + \left|-\mu,\,\tau\right>\right]$ and are made up of eigenstates of the operators corresponding to $p_{b}$ and $p_{c}$:
\begin{equation}
 \hat{p}_{b}\left|\mu,\,\tau\right>= \frac{1}{2}\gamma \mu \left|\mu,\,\tau\right>, \;\;\;\; \hat{p}_{c}\left|\mu,\,\tau\right>= \frac{1}{2}\gamma \tau \left|\mu,\,\tau\right>\:. \label{eq:peigen}
\end{equation}
The volume operator is also needed:
\begin{equation}
 V =\int d^{3}x\,\sqrt{|\mbox{det}E|} = 4\pi \sqrt{|p_{c}|} |p_{b}| \rightarrow \hat{V} = 4\pi \sqrt{|\hat{p}_{c}|} |\hat{p}_{b}| \:, \label{eq:volop}
\end{equation}
which is diagonal in the $\left|\mu,\,\tau\right>$ basis and possesses eigenvalues:
\begin{equation}
 V_{\mu\tau}=2\pi\gamma^{3/2}|\mu| \sqrt{|\tau|} \:. \label{eq:voleigs}
\end{equation}
As well, the co-triad operator can be created. In the notation of \cite{ref:AB}, $\omega_{c}:=\frac{\mbox{sgn}\,p_{c}\:|p_{b}|}{\sqrt{|p_{c}|}}$ and $\omega_{b}:=\mbox{sgn}p_{b} \sqrt{|p_{c}|}$. It is noted that the co-triad can be written in terms of the holonomy and volume: 
\begin{equation}
\omega_{c}=(2\pi\gamma)^{-1}\, \mbox{Tr}\left(\tau_{3} h_{R} \left\{h_{R}^{-1},\, V \right\}\right) \:, \nonumber
\end{equation}
where $h_{R}$ corresponds to the holonomy along an interval in the $R$-direction. The operator version is given by:
\begin{equation}
\hat{\omega}_{c}\left|\mu,\,\tau\right>=- i (2\pi\gamma)^{-1}\mbox{Tr}\left(\tau_{3} \hat{h}_{R} \left\{\hat{h}_{R}^{-1},\, \hat{V} \right\}\right)\left|\mu,\,\tau\right> =\frac{\sqrt{\gamma}}{2} |\mu| \left(\sqrt{|\tau+1|} - \sqrt{|\tau-1|}\right) \left|\mu,\,\tau\right> \:.
\end{equation}
 Similarly, for $\omega_{b}$ one can construct \cite{ref:AB}
\begin{equation}
\hat{\omega_{b}}\left|\mu,\,\tau\right> = \sqrt{\gamma}\, \mbox{sgn}(\mu) \sqrt{|\tau|} \left|\mu,\,\tau\right>
\end{equation} 

As is usual, a general state can be expanded in terms of the above eigenstates: $\left| \psi \right> = \underset{\mu\tau}\sum c_{\mu\tau}\left|\mu,\,\tau\right>$. 

Next the Hamiltonian constraint is constructed. One may pursue this in two ways. One way is to write the extrinsic  curvature connection, $K$ in (\ref{eq:qe3}), in terms of the modified connection $A$. The second way is to regard $K$ as the connection to be used. However, in this second case, the holonomies are to be constructed as functions of $K$, not $A$. It was shown in \cite{ref:AB} that utilizing the second method results in a Hamiltonian constraint of the form:
\begin{equation}
C_{Ham}=\frac{1}{\gamma^{2}} \frac{1}{\sqrt{\det(E)}} E^{ai}E^{bj} \epsilon_{ijk} \left[\gamma^{2} \Omega^{k}_{ab} - {^{o}F}^{k}_{\;ab}\right]\:.
\end{equation}
Here ${^{o}F}:=dK +\left[K,\,K\right]$ and $\Omega:=d\Gamma=-\sin\theta\,\tau_{3}d\theta\wedge d\varphi$ (which can be calculated utilizing the triad associated with the standard two-sphere metric). Written in terms of the co-triad one has:
\begin{equation}
\frac{1}{\sqrt{\det(E)}}\epsilon_{ijk}\tau^{i}E^{aj}E^{bk} =-\left(4\pi \gamma \mathcal{L}_{(k)}\right)^{-1} \epsilon^{abc}  \:{^{o}\omega}_{c}^{k}\, h^{\delta}_{k} \left\{h_{k}^{\delta-1}, \, V\right\}\:. \label{eq:firstham}
\end{equation}
Here, $h_{k}$ corresponds to the holonomy along an edge in the $k$-direction ($k=R,\, \theta, \, \phi$) (the index $k$ is summed over). $\mathcal{L}_{(R)}=  \mathcal{L}_{(\theta)} =\mathcal{L}_{(\varphi)}= \delta$, where $\delta$ is the length of the curve in the coordinate directions over which the holonomy is measured (superscripts $o$ denote that the quantity is in the $K$ connection representation). As well, the last part of the Hamiltonian constraint can be written in terms of the co-triad:
\begin{equation}
 {^{o}F}^{i}_{\;ab}\tau_{i} \approx \frac{{^{o}\omega}_{a}^{i}\:{^{o}\omega}_{b}^{j}}{\mathcal{A}_{(ij)}} \left(h_{(ij)}^{\mathcal{A}} -1\right) +\mathcal{O}(\delta)\:, 
\end{equation}
where $\mathcal{A}_{R\theta} = \mathcal{A}_{R\varphi}=\mathcal{A}_{\theta\varphi} = \delta^{2}$ and $h_{(ij)}:=h_{i}h_{j}h_{i}^{-1}h_{j}^{-1}\:$.

At this stage, all quantities are constructed out of holonomies (in ${^{o}\omega}$) and tetrads (in ${^{o}\omega}$ and $V$) and therefore we can go to the quantum picture by simply replacing these with their operator analogues. In order to make the constraint Hermitian, the \emph{gravitational constraint} is defined as $\hat{C}^{\delta}_{\mbox{\tiny{grav}}}:= \frac{1}{2} \left[\hat{C}^{\delta}+ \hat{C}^{\delta\; \dagger}\right]$. Without details, we quote the main results of this construction: The Hamiltonian constraint operator, acting on the states yields a \emph{difference} equation for $c_{\mu\tau}$:
\begin{align}
 \hat{C}^{\delta}_{\mbox{\tiny{grav}}}\,c_{\mbox{\tiny{$\mu,\tau$}}}=&2\delta \left(\sqrt{|\tau+2\delta|} +\sqrt{|\tau|}\right) \left[c_{\mbox{\tiny{$\mu+2\delta,\tau +2\delta$}}} - c_{\mbox{\tiny{$\mu-2\delta,\tau+2\delta$}}}\right] \nonumber \\
& +\left(\sqrt{|\tau+\delta|} - \sqrt{|\tau-\delta|}\right) \left[(\mu+2\delta) c_{\mbox{\tiny{$\mu+4\delta,\tau$}}} -(1+2\gamma^{2}\delta^{2})\mu\, c_{\mbox{\tiny{$\mu,\tau$}}} \right. \nonumber \\
& +\left. (\mu -2\delta)c_{\mu-4\delta, \tau} \right] +2\delta\left(\sqrt{|\tau-2\delta|} +\sqrt{|\tau|}\right) \left[c_{\mbox{\tiny{$\mu-2\delta,\tau-2\delta$}}} \right. \nonumber \\
&\left. - c_{\mbox{\tiny{$\mu+2\delta,\tau-2\delta$}}}\right]=0 \:. \label{eq:diffevolve}
\end{align}
To analyze the behavior at the singularity, one starts at some positive value of $\tau$ and utilizes (\ref{eq:diffevolve}) to evolve the $c_{\mu,\tau}$ to smaller values of $\tau$. The singularity resolution issue is insensitive to the choice of initial conditions (provided, of course, that they are not pathalogical).

It turns out that the coefficients are always regular throughout the evolution for all values of $\tau > 0$ as well as $\tau \leq 0$. From (\ref{eq:pcomps}) and (\ref{eq:peigen}), $\tau=0$ corresponds to the classical singularity. Therefore, the evolution of the $c_{\mu,\tau}$ coefficients is regular and may proceed \emph{beyond} the classical singularity. In this theory of quantum geometry then, the quantum analogue of $T=0$ is no longer a boundary of the space-time. In essence there is a smooth ``bounce'' which evolves to another large region of space-time. Expectation values of curvature encoding quantities are also finite. Interestingly, (though perhaps not surprising) mini-superspace reduced LQG cosmological models possess similar behavior where the big-bang is replaced by a smoothly evolving quantum bounce. (See \cite{ref:lqgcos1} and references therein.)

\section{Black Holes in Astrophysics}

Although there exists no unquestionable proof that black holes exist in nature, there is mounting evidence suggesting that black holes are present in our universe. The evidence must be, by definition, indirect. We overview here some of the observational evidence for the existence of black holes as well as how researchers utilize observations to deduce the properties of these fascinating objects.

Perhaps the first evidence that some extreme objects exist in the universe was the observation of X-ray sources outside the solar system and of quasars in the 1960s \cite{ref:quasars}. Quasars are objects which possess luminosities on the order of $10^{14}$ that of the sun. Matter accreting into a black hole could most easily explain such emissions of electromagnetic radiation. It is now also believed that black holes are related in some way to the observed gamma ray bursts. Since it is expected that almost all black holes have some amount of rotation, the Kerr solution provides a viable background in which to study these emissions. Astrophysical theory is suggestive that there is a limit on the angular momentum parameter of $-0.998M \leq a \leq 0.998M$ \cite{ref:thornelimit} with high rotations amost certainly containing an event horizon and therefore unlikely to be an alternative to a black hole without event horizon \cite{ref:cardspin}.

\begin{figure}[h!t]
\begin{center}
\includegraphics[bb=14 17 831 612, clip, scale=0.35, keepaspectratio=true]{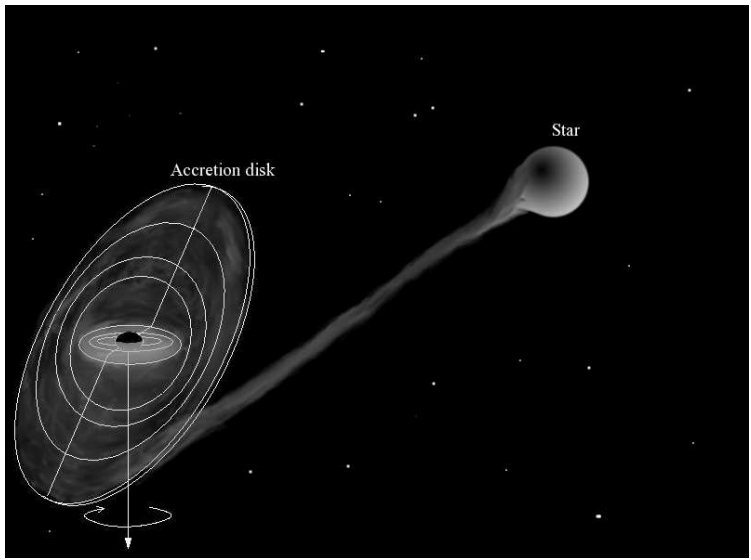}
\caption{{\small A rotating black hole accreting matter from a
nearby star. Although the outskirts of the accretion disk is tilted
with respect to the orbital plane, the inner regions are forced into
alignment with the orbital plane of the black hole.}}
\label{fig:accretiondisk}
\end{center}
\end{figure}

One of the earliest (and brightest) X-ray sources detected was Cygnus X-1 \cite{ref:cygx1}, \cite{ref:cygx1b}, which was noted to vary with time. A large number of X-ray sources have since been discovered, many of them associated with optically faint, distant stars. In such cases, it was not possible that the star itself could be emitting the X-rays. Instead, an argument was put forward that the X-rays originated from the accretion of the star's outer material onto a yet unseen companion object, likely a neutron star \cite{ref:accretemodel} \cite{ref:prendburb}. Further, it was postulated that the slightest amount of angular momentum in the material (likely inevitable) would preclude anything resembling radial in-fall and the material would, rather, be forced into a disk around the compact companion (see figure \ref{fig:accretiondisk}). Viscous drag forces would then heat up the material to high enough temperatures to emit the radiation observed. The Uhuru satellite confirmed that the stars indeed must be orbiting some companion object, which must be very compact. If the mass of the companion is above the neutron star limit (approximately 1.5 - 4 solar masses) then the likely alternative is a black hole. 

\begin{figure}[h!t]
\begin{center}
\includegraphics[bb=0 0 484 467, clip, scale=0.45, keepaspectratio=true]{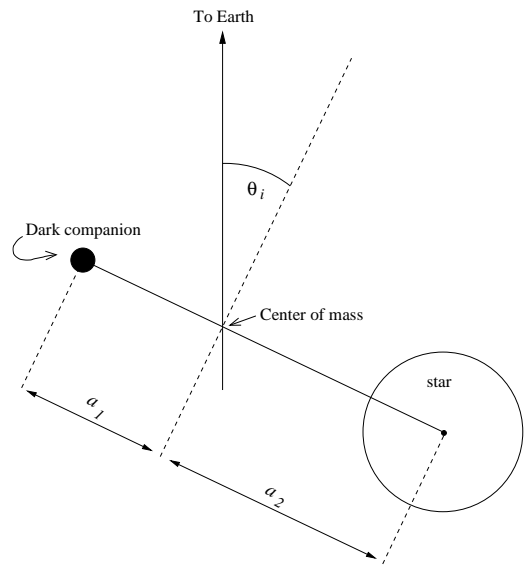}
\caption{{\small An example of a binary system.}} \label{fig:binary}
\end{center}
\end{figure}

The evidence for a binary system comes from the periodicity of the visible star's spectrum. One then needs to determine if the properties of the unseen companion allow it to be a neutron star or some more compact object. Consider a binary system as shown in figure \ref{fig:binary}.
As an approximation, Kepler's law can be utilized:
\begin{equation}
 T^{2}=\frac{M_{1} 4\pi^{2} a_{2}^{3}}{(M_{1}+M_{2})^{2}}\:, \label{eq:kepler}
\end{equation}
where $T$ is the orbital period and $M_{1}$ and $M_{2}$ represent the mass of the dark companion and observable star respectively. If the period, $M_{2}$, and $a_{2}$ are provided, the mass of the the companion may be determined. In practice it is usually difficult to determine $a_{2}$ as it requires knowledge of the inclination angle, $\theta_{i}$.

For the case of Cygnus X-1, the optical member is a hot OB star (HDE 226868) which typically have very large masses ($\gtrsim 20$ solar masses). With best estimates for the parameters, the conclusion is that the dark companion in this binary system possesses a mass of the order 10 solar masses, much greater than the neutron star limit. It is therefore likely that the companion is a black hole. One of the sources of uncertainty is the value of the mass of HDE 226868.

A class of X-ray objects are known as the X-ray transients. These objects emit X-rays periodically, followed by long periods of no emission. This allows one to study the optical companion in detail during the X-ray-quiet periods without the noise from the X-rays interfering with the observations. This is useful since detailed studies of the optical member of the binary will yield tighter constraints on the mass of the star and therefore on the mass of the companion object. In such a system, if the compact companion possessed a solid surface, it should be possible to see a characteristic emission of energy as the accreting material is brought to rest on the surface \cite{ref:solidsurf1}, \cite{ref:solidsurf2}, \cite{ref:narayan}. Observations of various sources seem to show no indication of such emission, and therefore the presence of an event horizon, as opposed to an object with a solid surface, is favored.

As mentioned earlier, the discovery of quasars in the 1960s has led to speculation that some compact object must be responsible for the emission of so much energy. Over the years the evidence has become compelling that the gravitational sources are likely super-massive black holes at the center of galaxies. These objects are now generally referred to as active galactic nuclei (AGN). These sources typically emit $10^{12} - 10^{14}$ solar luminosities and have length scales on the order of less than a light-year and in many cases the scale may be measured in light-hours. These scales are based on the fact that appreciable changes in a system can not occur on time-scales shorter than it takes light to cross the system.

\begin{figure}[h!t]
\begin{center}
\includegraphics[bb=1 83 304 314, clip, scale=0.70, keepaspectratio=true]{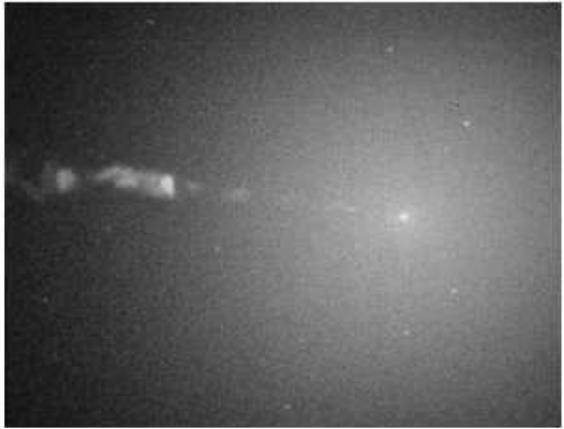}
\caption{{\small Jets of ionized gas being ejected from the galactic core of M87. Permission to use image, made available through NASA and STScI, kindly provided by Tod R. Lauer (National Optical Astronomy Observatory), Sandra M. Faber (UC Observatories/
Lick Observatory). (figure quality reduced for arXiv file size)}} \label{fig:agnjets}
\end{center}
\end{figure}

The physics involved in AGNs is similar to the compact binary described above. Namely, nearby matter is accreted into a disk around the black hole and X-rays are emitted via a friction mechanism. A natural question arises in the case of AGNs: Could the gravitational effects required to produce such X-ray emissions be due to the large number of stars and galactic matter near the core of the galaxy instead of a black hole? The constraints on the size, the lack of periodicities in the signals and the stability of the signals seem to favor a single central object (with masses of the order of $10^{10}$ solar masses!) instead of a widespread, non-uniform source \cite{ref:agnevid}. Also, there are ``jets'' of material present in may AGNs which remain aligned for time periods on the order of $10^{6}$ light-years (see figure \ref{fig:agnjets}). This indicates that a preferred axis must have been present in the system for at least that long, making a gravitationally bound compound object unlikely.

Modeling these jets is an extremely difficult task involving general relativistic magnetohydrodynamics. Large-scale computing must be employed in order to produce reliable results from the models. The jets can arise from a complex interplay between gas evaporating off of the accretion disk and magnetic fields present, known as the Blandford-Znajek process \cite{ref:BZproc}.

\begin{figure}[h!t]
\begin{center}
\includegraphics[bb=14 14 550 405, clip, scale=0.48, keepaspectratio=true]{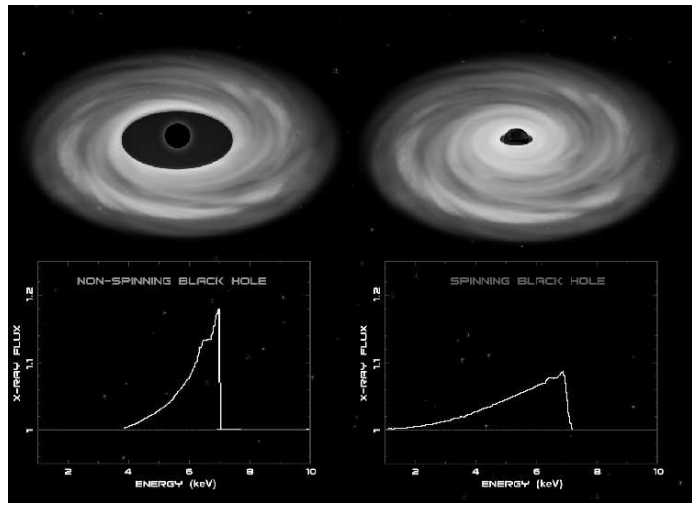}
\caption{{\small An example of an X-ray spectrum of iron atoms in
the vicinity of a non-spinning (left) and spinning (right) black
hole. A spinning black hole drags the particles around it, via the
Lense-Thirring effect, this allows for particles to orbit nearer to
the black hole. Kind permission to display this figure, from the
Harvard Chandra website, was granted by NASA/CXC/SAO.  }}
\label{fig:bhspin}
\end{center}
\end{figure}

One of the major sources of the X-rays is the iron $K\alpha$ line and there are several effects on this line. One effect is not strictly speaking gravitational in origin. It is the special relativistic doppler shift, where the line is blueshifted on one side of the accretion disk and redshifted on the other side. This will yield a doppler broadened line. Another effect is also present in the absence of gravity, this is the special relativistic beaming effect whereby the radiation intensity is amplified in the direction of particle motion compared to the intensity in the rest-frame. This effect is also differential in that the effect enhances the intensity on the approaching side of the disk.  Another effect is strictly gravitational in origin. This is the gravitational redshift and time dilation. Unlike the doppler shift, this shift does not depend on what side of the disk (approaching or receding side) the material is residing. Also, the gravitational time dilation has the effect of reducing the overall flux since the emitter is ``slowing down'' compared to an observer at infinity. These effects act to skew the line profile, as is illustrated in \ref{fig:bhspin}. The difference between the X-ray spectrum of matter accreting into a spinning versus non-spinning black hole is also displayed in this figure. This difference due to the spin arises from the fact that stable circular orbits in Schwarzschild geometry do not exist below $r=6M$, whereas for a rotating black hole this value is much smaller. Therefore, the gravitational redshift and time dilation can be much more pronounced in the case of a Kerr black hole as it is generally expected that the bulk of X-ray emission occurs in orbits at or above the stable orbit limit. With this assumption, data from the XMM-Newton satellite, analyzing the X-ray iron line of Seyfert galaxy MCG-6-30-15 constrains the Kerr spin parameter to be $|a|>0.93$ \cite{ref:xmm}. The rotational dragging of a Kerr black hole also has the effect of forcing the portion of the accretion disk that is close to the black hole to orbit in the equatorial plane (see figure \ref{fig:accretiondisk}).  

\begin{figure}[h!t]
\begin{center}
\includegraphics[bb=3 28 311 370, clip, scale=0.550, keepaspectratio=true]{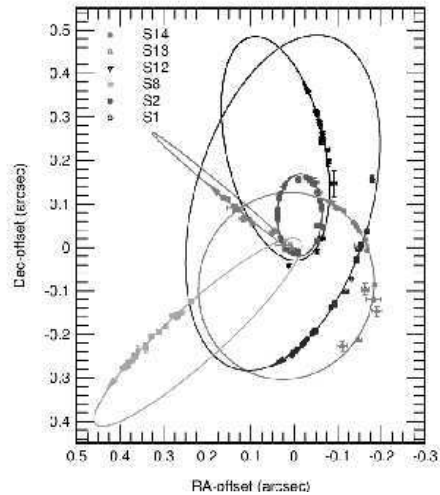}
\caption{{\small  Observation of the orbits of stars near the galactic center. This figure displays data points and best fit Keplerian orbits. Figure reproduced from \cite{ref:massgbh1} with kind permission from F. Eisenhauer and The Astrophysical Journal. (figure quality reduced for arXive file size)}} \label{fig:keporbs}
\end{center}
\end{figure}

Interestingly, there is strong evidence that there is a super-massive black hole at the center of our own galaxy (Sgr A$^*$). The ``close'' proximity of this black hole allows astronomers to directly measure its influence on its stellar neighbors \cite{ref:eckart} - \cite{ref:ghezsalim}. The optical range of frequencies cannot penetrate the galactic center so studies of the galactic black hole are usually performed utilizing the radio or the infra-red region. Knowing the orbital parameters between the black hole and its nearby stars, Kepler's law allows the determination of the approximate mass of the black hole. A mass of approximately $3.5\times 10^{6}$ solar masses is calculated for the mass of the galactic black hole \cite{ref:massgbh1}, \cite{ref:massgbh2}. Observational data relating to these orbits, along with Keplerian fits, are displayed in figure \ref{fig:keporbs}.

Other possible methods to detect black holes include: Gamma ray bursts, gravitational lensing (of background objects as well as of the orbital features of the accompanying binary star) \cite{ref:lens1}, \cite{ref:lens2}, and hopefully in the near future, gravitational wave signatures. Other proposals may be found in \cite{ref:salp}, \cite{ref:zelnov} and \cite{ref:zelnov2}.

We have only scratched the surface here regarding the observations and theoretical techniques used to study black holes in astrophysical contexts. There exist a number of excellent books and reviews on the subject. The interested reader is referred to the (much more thorough) review articles \cite{ref:celotti}, \cite{ref:galbhbook} and \cite{ref:muller} and the large number of references therein.

\section{Alternatives to Black Holes}
In this final section we will mention a few proposed alternatives to black holes along with possible measurements that may be performed in order to distinguish these objects from black holes. Some of these objects are black holes in the strict sense of the word. That is, they may contain a horizon but there is no singularity hiding behind it. It would be difficult, but not impossible, to distinguish some of these models from black holes \cite{ref:broder}. We list some of the alternatives here with a brief description. \vspace{0.1cm}

\noindent\textbf{Neutron stars with non-standard equation of state:} Perhaps the greatest Achille's heel to the arguments in favor of black holes as the likely candidate in the binary systems discussed above is the uncertainty in the equation of state. This is the main reason for the large uncertainty in the neutron star limit. The regime of neutron star density is above what can reliably be studied in a lab and therefore any properties at these densities are not well constrained. Since pressure is also a source of gravity in general relativity, modification of the equation of state could increase the maximum mass that neutron stars may possess and therefore some large mass objects thought to be black holes could turn out to be neutron stars. A general form of the equation of state was studied in detail by Rhoades and Ruffini \cite{ref:rr}. They made mild assumptions such as the speed of sound being bounded $0 \leq c_{s} \leq 1$, and that at lower densities (below some value $\rho_{0}$), it should produce equations of state thought to be well understood. In their study, they found a neutron star limit of approximately 3.2 solar masses. Adding rotation to the picture yields \cite{ref:friedip}:
\begin{equation}
 M_{\mbox{\tiny max}}\approx 8.4\left(\frac{\rho_{0}}{10^{14}}\right)^{-1/2} M_{\mbox{\tiny$\astrosun$}}\:.
\end{equation}
It is therefore not inconceivable that the neutron star limit could be as high as 8-10 solar masses or higher. 

A possible scenario is the ``Q-star'', which allows for the possibility of nucleon confinement under extreme conditions. In these theories, it is expected that under certain conditions, the equation of state differs strongly from standard ones, even at for relatively low densities (low values of $\rho_{0}$). Therefore, the assumptions used to derive the expression above are no longer valid. Such stars may possess masses on the order of 100 solar masses yet possess radii which are approximately 1.4 times the corresponding Schwarzschild radius \cite{ref:msn}. 

However, as mentioned previously, the absence of flare-ups due to material being brought to rest on a hard surface is in favor of a black hole instead of a neutron star or Q-star. As well, no reasonable value of $\rho_{0}$ would allow a neutron star scenario for the super-massive galactic black holes thought to be responsible for AGNs. \vspace{0.1cm}

\begin{figure}[h!t]
\begin{center}
\includegraphics[bb=0 0 353 222, clip, scale=0.75, keepaspectratio=true]{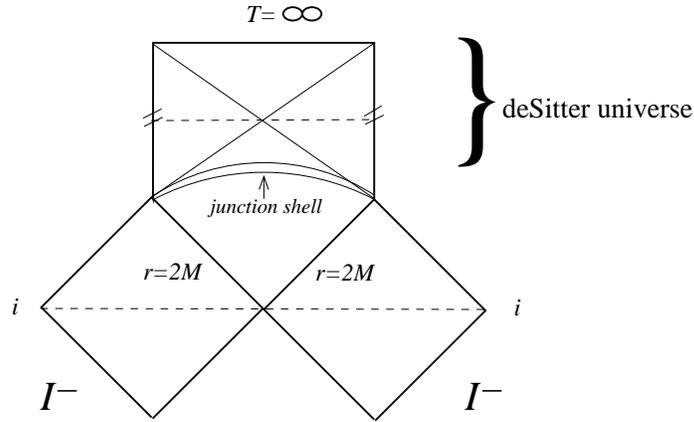}
\caption{{\small  A conformal diagram illustrating the space-time
that results from patching the $T$-domain of the Schwarzschild
space-time to a deSitter universe at a junction shell. Presumably a
phase-transition of the collapsing matter occurs at the shell
yielding the deSitter interior. $T$ denotes the interior time
coordinate whereas $r$ denotes the exterior radial coordinate.}}
\label{fig:schwdes}
\end{center}
\end{figure}

\noindent\textbf{Repulsive interiors:} This is not an alternative to a black hole as much as it is a possible alternative to the standard picture of an event horizon shielding a singularity. These scenarios basically stem from the fact that there is no reason to believe that the Schwarzschild solution is valid down to $r=0$. In the $T$-domain (the $r < 2M$ domain of the Schwarzschild solution) it is possible, for example, to patch the Schwarzschild solution to a deSitter metric via a shell located at some space-like surface. The idea is to preserve the properties of the event horizon, which seems to fit observational data, but modify the interior. This idea seems to date back to Sakharov and Gliner who considered the possibility that, under extreme conditions, matter would possess an equation of state of the form $\rho=-p$ \cite{ref:sakha}, \cite{ref:gliner}. Explicit constructions of this model were performed in \cite{ref:fmm} and alternates of this model were also considered in \cite{ref:dym1}-\cite{ref:dym3}. A conformal diagram of Schwarzschild space-time with a deSitter interior is shown in figure \ref{fig:schwdes}.

\noindent\textbf{Gravastars:} A recent extension of the above idea is the gravitational vacuum star, or gravastar. The gravastar idea originated with P. Mazur and E. Mottola as an alternative to a black hole and possesses \emph{no} event horizon \cite{ref:MM1} - \cite{ref:MM3}. In the gravastar picture, quantum vacuum fluctuations are expected to
play a non-trivial role in the collapse dynamics. A phase
transition is believed to occur yielding a repulsive de Sitter
core which aids in balancing the collapsing object and thus preventing
horizon (and singularity) formation. This transition is expected to occur very
close to the limit $2m(r)/r=1$ so that, to an outside observer,
it would be very difficult to distinguish the gravastar from a true black hole. The idea of a phase transition of the vacuum from a $\Lambda \approx 0$ state to a non-negligible $\Lambda$ state is motivated from the behavior of Bose-Einstein condensates. The final gravastar configuration would also possess much less entropy than a black hole of similar size and therefore the problem of where the enormous black hole entropy comes from is alleviated.

\begin{figure}[ht]
\begin{center} \includegraphics[bb=0 0 374 180, scale=0.9, clip, keepaspectratio=true]{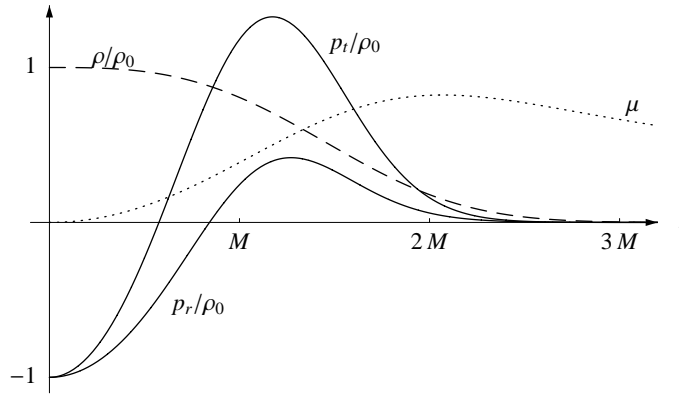}
\caption{{\small The gravastar with energy density profile
$\rho=\rho_0\,\exp[-(r/r_0)^n]$ and anisotropy
$\tilde{\Delta}=\alpha^{2}\;(\rho/\rho_{0})\;\mu/ 12$. Displayed
are: radial (lower solid line) and transversal (upper solid line)
pressures, energy density (dashed line) and the compactness (dotted
line). In this example the parameters are $n=3$, total mass of
configuration $M=1$ and maximal compactness within the gravastar is
$\mu_{\mbox{max}}=0.80$. Notation is as follows:
$\tilde{\Delta}:=\frac{p_t-p_r}{\rho_{0}}=\frac{\alpha^2}{12}\frac{2m(r)}{r}\frac{\rho}{\rho_{0}}$,
with $\alpha$ and $\rho_{0}$ constants. $\mu$ is the ``compactness''
function $2m(r)/r$. The transverse pressure, radial pressure and
energy density are denoted as $p_{t}$, $p_{r}$ and $\rho$
respectively.}} \label{fig:gravapres}
\end{center}
\end{figure}

The original Mazur - Mottola model consisted of a deSitter interior separated from a Schwarzschild exterior via a finite shell with an equation of state satisfying $\rho=+p$ (with thin shells on either side for patching purposes). It was later shown that, were a transition between a deSitter center and Schwarzschild exterior to be smooth and yield physically reasonable outer layers, anisotropic pressures must be present within the structure \cite{ref:visgrava}. Models with continuous pressures satisfying various equations of state were explicitly constructed in \cite{ref:dubgrava}. Examples of the pressure and density profiles for a sample gravastar (originally displayed in \cite{ref:dubgrava}) are displayed in figure \ref{fig:gravapres}. Lobo and Arellano have studied several variants of gravastars or gravastar-like objects in \cite{ref:lobo1} \cite{ref:lobo2}.

Note that by definition, the strong energy condition cannot be satisfied in any model with a deSitter region. A thorough discussion of gravastar energy conditions may be found in \cite{ref:sachecond}.

In the case of anisotropic models, one way to distinguish their presence from either a black hole or a neutron star is via their surface redshift. It is known (see \cite{ref:boehm} and references therein) that stability in anisotropic spheres allows for a higher maximum redshift than for a stable perfect fluid sphere of similar mass, due to increased allowable compactness. Also, Chirenti and Rezzolla have discussed how to distinguish a gravastar from a black hole via quasi-normal mode analysis \cite{ref:distinguish}.

The list presented above is not exhaustive but should cover some of the most popular alternatives to black hoes. Another interesting alternative for example, put forward by Robertson and Leiter, is the magnetic eternally collapsing object (or MECO) (see references \cite{ref:meco1}, \cite{ref:meco2}, \cite{ref:meco3} and references therein.)

\section*{Acknowledgments}
I am grateful to S.~Kloster and J.~Br\"{a}nnlund for discussions and help with improving the manuscript. I thank M.~Bojowald for kindly clarifying an issue in the quantum singularity resolution. I would also like to thank the various authors, journals and institutions which have granted permission to use their figures in this paper (acknowledged individually in the figure captions).


\label{lastpage-01}

\end{document}